\documentclass[10pt,twocolumn,showpacs,preprintnumbers,amsmath,amssymb,aps,prb,longbibliography,superscriptaddress]{revtex4-2}
\usepackage{mathrsfs}
\usepackage{graphicx}
\usepackage{dcolumn}
\usepackage{bm}
\usepackage{amsmath}
\usepackage{amsfonts}
\usepackage{color}
\usepackage{hyperref}

\usepackage{cleveref}
\crefname{appendix}{App.}{Apps.}
\crefname{equation}{Eq.}{Eqs.}
\crefname{figure}{Fig.}{Figs.}
\crefname{table}{Tab.}{Tabs.}
\crefname{section}{Sec.}{Secs.}

\begin{document}

%\title{A Quantum Model of Geiger Counter (Avalanche)}
\title{A Quantum Breakdown Model: from Many-body Localization to Chaos with Scars}
%Quantum Chaos with Dark Zero Modes and 

\author{Biao Lian}
\affiliation{Department of Physics, Princeton University, Princeton, New Jersey 08544, USA}

\begin{abstract}
We propose a quantum model of fermions simulating the electrical breakdown process of dielectrics. The model consists of $M$ sites with $N$ fermion modes per site, and has a conserved charge $Q$. It has an on-site chemical potential $\mu$ with disorder $W$, and an interaction of strength $J$ restricting each fermion to excite two more fermions when moving forward by one site. We show the $N=3$ model with disorder $W=0$ show a Hilbert space fragmentation and is exactly solvable except for very few Krylov subspaces. The analytical solution shows that the  $N=3$ model exhibits many-body localization (MBL) as $M\rightarrow\infty$, which is stable against $W>0$ as our exact diagonalization (ED) shows. At $N>3$, our  ED suggests a MBL to quantum chaos crossover at small $W$ as $M/N$ decreases across $1$, and persistent MBL at large $W$. At $W=0$, an exactly solvable many-body scar flat band exists in many charge $Q$ sectors, which has a nonzero measure in the thermodynamic limit. We further calculate the time evolution of a fermion added to the particle vacuum, which shows a breakdown (dielectric) phase when $\mu/J<1/2$ ($\mu/J>1/2$) if $W\ll J$, and no breakdown if $W\gg J$.
%The breakdown is local when $M/N\gg1$, and is global when $M/N\ll 1$.
\end{abstract}

\date{\today}

\maketitle

\section{Introduction}

The breakdown of dielectrics in an electric field (the Townsend avalanche) \cite{townsend1911} is one of the most violent far-from-equilibrium systems. Generically, it involves a spatially asymmetric process which creates more and more particles towards a fixed spatial direction, as driven by certain forces (electric field, enthalpy gradient, etc). As a result, a microscopic perturbation can be amplified into a macroscopic signal in a short time, producing a particle avalanche. Therefore, the breakdown process is often utilized in designing quantum measurement apparatuses, such as the Geiger Counter which works near the breakdown electric field of inert gases \cite{rutherford1908,geiger1928}. Similar processes also occur in chemical reactions and generic chain reactions. However, the study of the breakdown phenomenon has been largely restricted to phenomenological and semiclassical models, since the quantum problem inevitably involves an enormous Hilbert space and strong many-body interaction.

The studies in the past decades have significantly advanced our knowledge on non-equilibrium quantum dynamics of many-body systems. The thermalization is revealed to be closely related to many-body quantum chaos by the eigenstate thermalization hypothesis \cite{jensen1985,deutsch1991,srednicki1994,dalessio2016}, and solvable quantum chaotic models, particularly the Sachdev-Ye-Kitaev (SYK) models \cite{sachdev1992fk,polchinski2016xgd,maldacena2016hyu,kitaev2017awl}, manifest the quantum butterfly effect of perturbation amplifications. On the other hand, quantum systems lacking thermalization have been investigated extensively, which are known as the many-body localization (MBL) phase \cite{basko2006,gornyi2005,oganesyan2007,marko2008,pal2010}. The MBL has also been numerically studied in interacting fermions in an electric field (interacting Wannier-Stark) \cite{Nieuwenburg_2019,schulz2019} and similar models \cite{Moudgalya_2021,herviou2021,sala2020,khemani2020}. In addition, non-Hermitian effective quantum models have been proposed to describe the electrical breakdown of the Mott insulator \cite{fukui1998,oka2010}. However, the Hermitian quantum modeling of the electrical breakdown phenomena has not yet been explored. Models showing Hilbert space fragmentations have also been studied as systems lacking thermalization  \cite{Moudgalya_2021,herviou2021,sala2020,khemani2020,znidaric2013,yangzc2020,moudgalya2020}. In addition, many-body quantum scars have been studied as non-thermal eigenstates in quantum chaotic systems \cite{bernien2017,moudgalya2018,schecter2018,turner2018a,turner2018b,choi2019scar,ho2019,bull2019,lin2019,khemani2019,scherg2021,kao2021,jepsen2022,su2022scar,desaules2022,desaules2022b}. These developments have provided new tools for us to study the quantum physics of the electrical breakdown.

In this paper, we introduce a \emph{quantum breakdown model} with a \emph{spatially asymmetric} interaction, to give a simplified quantum modeling of the breakdown process. The model consists of a 1D chain of $M$ sites, and each site consists of $N$ fermion degrees of freedom. Except for an on-site chemical potential, the model only contains an asymmetric nearest-neighbor interaction that annihilates (creates) a fermion on the $m$-th site and creates (annihilates) three fermions on the $(m+1)$-th site ($1\le m<M$). The model is to provide a simplified model for the dielectric gas breakdown in an electric field (the Townsend avalanche \cite{townsend1911}), where each site represents a layer of the gas perpendicular to the electric field, and the layer separation is roughly the mean free path. When an atom (or molecule) in a certain layer is ionized into an ion and electron, the electron will be accelerated by the electric field and excite more ionizations (two more in the model here) in the next layer through collisions, as shown in Fig. \ref{fig-model}(a). Therefore, although the microscopic interaction in the gas is generically inversion symmetric, the effective interaction in the electric field is inversion asymmetric. 

We note that our model here is an oversimplified model for breakdown in the following sense: first, the electric field (or other generalized force) is not explicitly present, instead its effect is integrated into the asymmetric interaction. Secondly, we have ignored the spatial structure within each layer of gas, assuming an electron in one layer can excite any electrons in the next layer. Despite these simplifications, we can approximately identify the asymmetric interaction strength $J$ as the voltage difference between adjacent layers, and view the chemical potential $\mu$ in each layer as the ionization energy of an atom. We also include a disorder strength $W$ giving fluctuations in $\mu$.

The quantum breakdown model exhibits a surprisingly rich physics. The model has a conserved charge $Q$, which significantly reduces its numerical complexity. We first show that the $N=3$ model with disorder strength $W=0$ shows a Hilbert space fragmentation into exponentially large number of Krylov subspaces in all charge $Q$ sectors, and is exactly solvable except for very few Krylov subspaces of certain charge $Q$ sectors. The exact solution shows the $N=3$ model is in a 1D MBL phase in the $M\rightarrow\infty$ limit (the 1D limit), and we further verified the robustness of the MBL under nonzero disorder $W>0$ via exact diagonalization (ED). For $N>3$, we perform ED calculations. At weak disorder strength $W$, from the level spacing statistics (LSS) and eigenstate entanglement entropies, we find the model shows MBL features when $M/N>1$, while is quantum chaotic when $M/N<1$. This lead us to the conjecture that the model at small $W$ is in a 1D MBL phase with localization length of order $N$ in the limit $M/N\rightarrow\infty$, and is quantum chaotic in the limit $M/N\rightarrow 0$ (the 3D limit), with a crossover around $M/N\approx 1$. At sufficiently large $W$, the model shows MBL features irrespective of $M/N$. Moreover, when $W=0$, in many charge $Q$ sectors, there exists a set of exactly solvable degenerate eigenstates forming a many-body scar flat band, leading to quantum scar states with a nonzero measure in the thermodynamic limit.

We further investigate the time evolution of a fermion added to the first layer of the model, to examine the breakdown phenomena. With weak disorders $W\ll J$, irrespective of $M/N$, our calculations show the system is in a breakdown phase when $\mu/J<1/2$  ($\mu\ge0$), and in a dielectric phase (no breakdown) when $\mu/J>1/2$. The ratio $M/N$ affects the spatial ranges of the breakdown. When $M/N>1$, the model can only have a local breakdown with the particles ionized within order $N$ number of sites after the initial perturbation. When $M/N<1$, the model will exhibit a global breakdown with particles ionized in the entire system. With strong disorders $W\gg J$, the breakdown is absent because of the persistent MBL for any ratio $M/N$.

The rest of this paper is organized as follows. The quantum breakdown model and its conserved charge is defined in \cref{sec:model}. In \cref{sec-N=3-W=0,sec-N=3-W>0}, we first give the exact and almost exact solution of the $N=3$ model at disorder $W=0$, and then compare with the ED results at both $W=0$ and $W>0$. \cref{sec:N>3} shows the ED calculations for $N>3$ models suggesting the MBL to chaos crossover around $M/N\approx1$ at small $W$, the persistent MBL at large $W$, and gives the theoretical understanding of the many-body scar flat bands. \cref{sec-TE} presents the time-evolution of a fermion added to the particle vacuum of the model, which reveals the breakdown transition at $\mu/J=1/2$ at small $W$, and absence of breakdown at large $W$. Lastly, remarks and future directions are discussed in \cref{sec:discussion}.

\section{The quantum breakdown model}\label{sec:model}

\subsection{The Hamiltonian}

\begin{figure}[tbp]
\begin{center}
\includegraphics[width=3in]{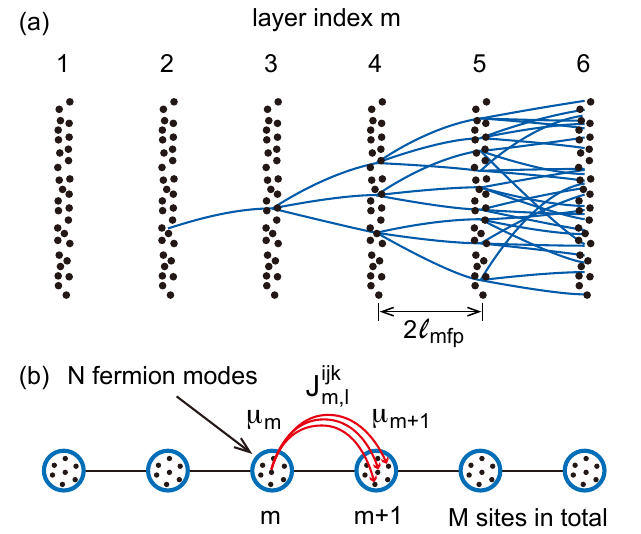}
\end{center}
\caption{(a) Simplified illustration of the electrical breakdown, where an electron in a layer hits the next layer and creates two more electrons, leading to a particle avalanche. (b) The quantum breakdown model in a 1D lattice with $M$ sites (open boundary), and $N$ fermion modes per site. The only interaction annihilates a fermion in site $m$ and creates three fermions in site $m+1$ (and the inverse process).
}
\label{fig-model}
\end{figure}

Fig. \ref{fig-model}(a) gives a simplified illustration the breakdown (Townsend avalanche) process of a dielectric gas in a large enough electric field (to the left), where we assume the gas consists of neutral atoms (the black dots). When an atom is ionized into an electron and an ion, the electron will be accelerated in the direction opposite to the electric field, and then hit and ionize more atoms, which will eventually lead to a particle Townsend avalanche as illustrated by the blue lines. In principle, the ions can induce a similar particle avalanche, but it would happen in a much longer time scale because of their much heavier mass, and may be terminated by the recombinations of ions with electrons from the electrode on the left. Moreover, if the dielectric is a solid instead of a gas, the ions will stay static. Therefore, we will ignore the ions and focus on the electrons to simplify the quantum model of the breakdown.

By ignoring ions, we can assume each ionized electron in layer $m$ can hit layer $m+1$ and create more ionized electrons via interaction. The cost of ignoring the ions is that one can only write down interactions with electron number changing by an even number, since electrons are fermions. Therefore, the simplest breakdown interaction is an inversion asymmetric interaction annihilating one electron in layer $m$ and creating three electrons in layer $m+1$ (and its Hermitian conjugate). The distance between two adjacent layers is thus approximately $2\ell_{\text{mfp}}$ for 2 collisions to happen, where $\ell_{\text{mfp}}$ is the mean free path of the electrons.

To further simplify the model, we ignore the spatial structure within each layer, and treat each layer as a site of a 1D lattice with an open boundary condition, as shown in Fig. \ref{fig-model}(b). By assuming $M$ sites (layers) in total and $N$ complex fermion (electron) modes in each site, we define the quantum breakdown model Hamiltonian in the 1D lattice as containing three parts:
\begin{equation}\label{eq:model}
H=H_I+H_\mu+H_{\text{dis}}\ .
\end{equation}

The first part is the asymmetric breakdown interaction taking the form
\begin{equation}\label{eq-HI}
H_I=\sum_{m=1}^{M-1} \sum_{i<j<k}^{N} \sum_{l=1}^{N} \left(J_{m,l}^{ijk} c^\dag_{m+1,i} c^\dag_{m+1,j} c^\dag_{m+1,k} c_{m,l} + h.c.\right),
\end{equation}
where $c_{m,i}$ and $c_{m,i}^\dag$ are the annihilation and creation operators of the $i$-th fermion mode on the $m$-th site ($1\le i\le N$, $1\le m\le M$), and $J_{m,l}^{ijk}$ are complex parameters which are totally antisymmetric in indices $i,j,k$. The notation $h.c.$ stands for Hermitian conjugate of the first part, so $H_I$ is Hermitian. We shall consider the cases with both uniform and random interaction $J_{m,l}^{ijk}$. For $H_I$ not to vanish, we require the number of fermion modes per site $N\ge3$. 
%As a result, one expects an exponential growth in the number of excited fermions in an avalanche event, as illustrated in Fig. \ref{fig-model}(a).

The second part of the Hamiltonian is an on-site chemical potential uniform within each site:
\begin{equation}\label{eq:number-operator}
H_\mu=\sum_{m=1}^{M}\mu_m \hat{n}_m\ ,\qquad \hat{n}_m=\sum_{i=1}^{N}c^\dag_{m,i}c_{m,i}\ ,
%H_0=\sum_{m=1}^{M}\sum_i^{N} \mu_m c^\dag_{m,i}c_{m,i}=\sum_{m=1}^{M}\mu_m \hat{n}_m\ ,
\end{equation}
where $\hat{n}_m$ is the fermion number operator on site $m$, and to be general, we assume the chemical potential $\mu_m$ can depend on site $m$.
%\begin{equation}
%\hat{n}_m=\sum_{i=1}^{N}c^\dag_{m,i}c_{m,i}
%\end{equation}

%where $m$ is the layer index, $\mu_m$ can be understood as the ionization energy in layer $m$. There are $N$ fermion modes (with annihilation operators $c_{m,i}$) in layer $m$, and we have defined

The last part is a disorder potential within each site:
\begin{equation}\label{eq:model-Hd}
H_{\text{dis}}=\sum_{m=1}^{M}\sum_i^{N} \nu_{m,i}c^\dag_{m,i}c_{m,i}\ ,
\end{equation}
where $\nu_{m,i}$ are Gaussian random potentials with mean value and variance given by
\begin{equation}\label{eq-W}
\langle \nu_{m,i}\rangle=0\ ,\qquad \langle \nu_{m,i}^2\rangle =W^2\ .
\end{equation}
The constant $W\ge0$ characterizes the disorder potential strength. We note that one could add a non-diagonal random hopping term $\nu_{m,ij}c^\dag_{m,i}c_{m,j}$ within a site, but one could always rewrite the model in the diagonal fermion basis of the hopping matrix $\nu_{m,ij}$ on each site $m$, after which the disorder potential will take the diagonal form of Eq. (\ref{eq:model-Hd}) without loss of generality.

For later convenience, we define the interaction strength between sites $m$ and $m+1$ as
\begin{equation}\label{eq-Jm}
J_m=\left(\frac{1}{N}\sum_{i<j<k}^{N} \sum_{l=1}^{N} \left|J_{m,l}^{ijk}\right|^2\right)^{1/2}\ .
\end{equation}
%namely, $J_m$ is the root mean square of $|J_{m,l}^{ijk}|$ over all the fermion mode indices $i,j,k,l$. 
In particular, when the system is on average uniform, which is the case we will focus on dominantly, we can assume both the chemical potential and the interaction strength are site $m$ independent:
\begin{equation}\label{eq-muJ}
\mu_m=\mu\ ,\qquad J_m=J\ .
\end{equation} 
In the case of fully random interactions $J_{m,l}^{ijk}$, we assume the interactions are generated from a Gaussian distribution with statistical mean and variance
\begin{equation}\label{eq-J}
\langle J_{m,l}^{ijk}\rangle=0\ ,\quad \langle\left|J_{m,l}^{ijk}\right|^2\rangle=\frac{3! J^2}{N(N-1)(N-2)}\ ,
\end{equation}
where $J$ is defined in consistency with Eq. (\ref{eq-muJ}). We note the similarity of such an interaction to that in the SYK model \cite{sachdev1992fk,polchinski2016xgd,maldacena2016hyu,kitaev2017awl}.

\subsection{Conserved charge}

With the open boundary condition as shown in Fig. \ref{fig-model}(b), the quantum breakdown model in Eq. (\ref{eq:model}) has a conserved global U(1) charge
\begin{equation}\label{eq:charge}
Q=\sum_{m=1}^M 3^{M-m} \hat{n}_m\ .
\end{equation}
In other words, each fermion in layer $m$ carries an effective charge $q_m=3^{M-m}$. The breakdown process can thus be understood as charged particles keeping splitting into partons carrying a $1/3$ factor smaller charge as the layer index $m$ increases by one. This allows us to study a fixed charge $Q$ sector each time, which has a much smaller Hilbert space dimension than that of the entire model.

We note that one could complicate the Hamiltonian of Eq. (\ref{eq:model}) by adding other terms, such as the hopping between different sites. This will, however, lead to the loss of the conserved charge $Q$, making the analytical and numerical studies much more difficult. In this paper, we restrict ourselves to the model in Eq. (\ref{eq:model}) with the conserved charge $Q$.

\subsection{Chiral symmetry}\label{sec:chiral}

We can define a unitary chiral transformation $C$ as
\begin{equation}\label{eq:chiral}
C (c_{m,l},c_{m,l}^\dag) C^{-1}=(-1)^m (c_{m,l},c_{m,l}^\dag)\ ,
\end{equation}
and it is straightforward to see that the model Hamiltonian $H=H_I+H_\mu+H_{\text{dis}}$ transforms as
\begin{equation}
C (H_I+H_\mu+H_{\text{dis}}) C^{-1}=-H_I+H_\mu+H_{\text{dis}}\ .
\end{equation}
Moreover, $[C,Q]=0$. Therefore, when the disorder potential $W=0$, namely $H_{\text{dis}}=0$, the Hamiltonian $H$ at chemical potentials $\mu$ and $-\mu$ are exactly opposite to each other via a chiral transformation, and thus have exactly opposite energy spectra in the same charge $Q$ sector. We call this property the \emph{chiral symmetry} between $\mu$ and $-\mu$. When $W>0$, the chiral symmetry still approximately true due to randomness of $H_{\text{dis}}$. Therefore, we will only focus on $\mu\ge 0$ hereafter. 

\subsection{Physical meaning of the parameters}\label{sec:model-physical}

In a realistic breakdown system, both the number of fermion modes $N$ per layer and the number of layers $M$ are large. We can view the ratio $M/N$ as proportional to the ratio of the length (in the direction of electric field) and cross sectional area of the space containing the dielectric gas. The ratio $M/N$ is thus tunable by the geometry of the container. In the 1D limit where the container is a quasi-1D long tube with a small cross sectional area, we have $M/N\gg1$. In contrast, in the 3D limit where the container has similar sizes in all the three directions, we have $M/N\ll 1$.
%As we will show, the quantum breakdown model in Eq. (\ref{eq:model}) exhibits distinct behaviors in energy spectrum and eigenstates with respect to $N/M$. When $M/N>1$, the model shows features of the many-body localization (MBL) phase, and exhibits a fractal-like many-body energy spectrum. When $M/N<1$, the model shows many-body quantum chaos with quantum scar states.

Physically, the interaction strength $J$ in Eq. (\ref{eq-muJ}) is approximately the average energy a fermion can gain by turning into three fermions in the next site. Therefore, one can identify $J$ with the electric potential energy difference $2eE_{\text{ele}}\ell_{\text{mfp}}$ of an ionized electron between neighboring layers, where $e$ is the electron charge, and $E_{\text{ele}}$ is the electric field. The chemical potential $\mu\ge0$ can be understood as the ionization energy for creating an electron, with a disorder strength $W$ due to fluctuations, which will disfavor breakdown. One thus expects the quantum dynamics of the system to undergo a dielectric breakdown transition as the ratio $\mu/J$ decreases towards zero, which is studied in Sec. \ref{sec-TE}.

%In most of the discussion in this paper, we will assume $N_m=N$ is independent of layer $m$. 

\section{The $N=3$ model with $W=0$}\label{sec-N=3-W=0}

We first study the model with $N=3$, the minimal number of fermion modes per site for the interaction in Eq. (\ref{eq-HI}) not to vanish. In this section, we fix the disorder potential strength $W=0$ (defined in Eq. (\ref{eq-W})), namely, $H_{\text{dis}}=0$. We will show that this $N=3$ model with $W=0$ shows a Hilbert space fragmentation into an exponentially large number of Krylov subspaces in each charge $Q$ sector, and is exactly solvable except for very few Krylov subspaces in certain charge $Q$ sectors, regardless of the parameters $\mu_m$ and $J_{m,l}^{ijk}$. The resulting many-body eigenstates indicate that the model is in a 1D MBL phase as $M\rightarrow\infty$, as summarized in \cref{fig-MBL}, and the energy spectrum exhibits an interesting fractal-like structure. The $N=3$ model with disorder strength $W>0$ will be studied in Sec. \ref{sec-N=3-W>0}, where we find the MBL phase remains robust.

\subsection{Model simplification}

With disorder potential $W=0$, we are allowed to simplify the model interactions by U(3) unitary rotations of the fermion basis $c_{m,l}$ in each site. We define a new fermion basis $f_{m,l}$ by
\begin{equation}\label{eq:trans-H3}
c_{m,l}=\sum_{l'=1}^3 U_{ll'}^{(m)}f_{m,l'}\ , 
\end{equation}
where the unitary matrices $U^{(m)}$ satisfy $\sum_{l}J_{m,l}^{123}U_{ll'}^{(m)}= \sqrt{3}J_m\delta_{1,l'} \det \left(U^{(m+1)}\right)$, 
with the interaction strength $J_m$ defined in Eq. (\ref{eq-Jm}), and $\delta_{l,l'}$ is the Kronecker delta function. 
%\begin{equation}\label{eq:Jm-N=3}
%J_m=\sqrt{\sum_{l=1}^3\left|J_{m,l}^{123}\right|^2}\ ,\qquad (1\le m\le M-1)\ .
%\end{equation}
Note that the product operator in each site $c_{m,1}c_{m,2}c_{m,3}=\det \left(U^{(m)}\right)f_{m,1}f_{m,2}f_{m,3}$ is invariant up to a phase factor $\det \left(U^{(m)}\right)$. This transforms the total Hamiltonian in Eq. (\ref{eq:model}) into the $f_{m,l}$ basis:
\begin{equation}\label{eq:H3}
\begin{split}
&H=H_I+H_\mu\ ,\\
&H_I=\sum_{m=1}^{M-1} \Big[\sqrt{3}J_m \Big(\prod_{j=1}^{3}f^\dag_{m+1,j}\Big)  f_{m,1} + h.c.\Big],\\
&H_\mu=\sum_{m=1}^{M}\mu_m \hat{n}_m\ ,
\end{split}
\end{equation}
where $\hat{n}_m=\sum_{l=1}^3 f_{m,l}^\dag f_{m,l}$ is equivalent to the definition in Eq. (\ref{eq:number-operator}). Note that only the first fermion mode $f_{m,1}$ in each site can excite three fermions in the next site.

%We now show that the above $N_m=3$ model in Eq. (\ref{eq:H3}) is almost analytically solvable. 

\subsection{Eigenstates of the $Q=3^{M-1}$ charge sector}\label{sec:sectorQ1-N=3}

The Hamiltonian is block diagonal in each charge $Q$ sector, the Hilbert space of which is spanned by the basis of all Fock states with integer fermion number $n_m=\langle \hat{n}_m\rangle$ on each site $m$, satisfying $Q=\sum_{m} 3^{M-m} n_m$.

%We can consider a fixed conserved charge sector $Q$ as defined in Eq. (\ref{eq:charge}), which yields a constraint on the fermion number configurations $\{n_m\}$ in each layer. Here $n_m=\langle \hat{n}_m\rangle$ is the site $m$ fermion number of a Fock eigenstate of the particle number operator $\hat{n}_m$ in \cref{eq:number-operator}.

As a prototypical example, we investigate the $Q=3^{M-1}$ charge sector in this subsection. This is the charge sector of adding one fermion to the first site of the particle vacuum state $|0\rangle$ (vacuum state defined by $\hat{n}_m|0\rangle=0$ for all $m$). The fermion numbers of the Fock state basis are thus limited to the following string configurations:
\begin{equation}\label{eq-config-N=3}
n_m=2\vartheta_{m_s-m+1}-2\delta_{m,1}+\delta_{m,m_s}\ ,
\end{equation}
where $m_s$ is an integer satifying $1\le m_s\le M$, and we have defined the integer variable Heaviside function $\vartheta_m$ as %$\eta_m=1$ if $m>0$, and $\eta_m=0$ if $m\le 0$ ($m\in\mathbb{Z}$). 
\begin{equation}\label{eq-heaviside}
\vartheta_m=
\begin{cases}
&1\ ,\quad (m>0) \\
&0\ ,\quad (m\le 0)
\end{cases}
\ ,\qquad (m\in\mathbb{Z}).
\end{equation}
%\begin{equation}\label{eq-config-N=3}
%n_m=
%\begin{cases}
%& 2-2\delta_{m,1}+\delta_{m,m_s}\ , \quad (m\le m_s) \\
%& 0\ ,\qquad\qquad\qquad\quad \qquad (m>m_s)
%\end{cases}
%\end{equation}
More explicitly, the fermion number configurations in \cref{eq-config-N=3} takes the form $\{n_m\}=\{0,2,\cdots,2,3,0,\cdots,0\}$ when $m_s>1$, where the $m_s$-th site has $3$ fermions and is the last site with a nonzero fermion number (i.e., the site the fermions have spread to). When $m_s=1$, one has $\{n_m\}=\{1,0,\cdots,0\}$, and we call this configuration the \emph{reference configuration} of charge $Q=3^{M-1}$, which is the configuration with the \emph{smallest} total number of fermions possible in this charge $Q$ sector.

The Fock states with $n_m$ given in Eq. (\ref{eq-config-N=3}) with all possible $m_s$ constitute the Hilbert space basis of the charge $Q=3^{M-1}$ sector, which has a Hilbert space dimension $h_{Q}=\frac{3^{M-1}+5}{2}$. In the below, we will show that this Hilbert space further fragments into an exponentially large number of disjoint Krylov subspaces, each of which has its Hamiltonian equivalent to a single-particle tight-binding model.

\begin{figure}[tbp]
\begin{center}
\includegraphics[width=3.3in]{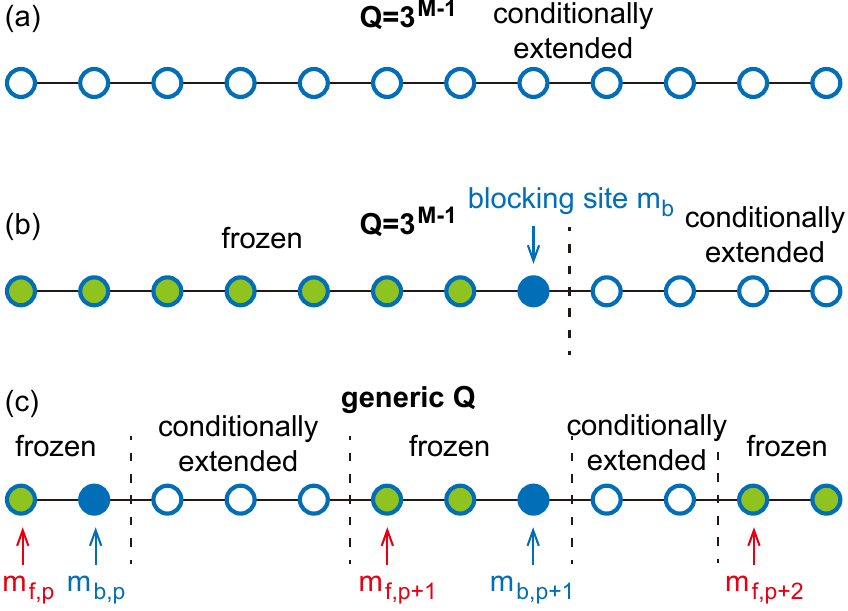}
\end{center}
\caption{Illustration of the eigenstates of the $N=3$ model $W=0$. (a) The Krylov subspace without blocking in the charge $Q=3^{M-1}$ sector, which is conditionally extended. (b) Krylov subspaces with blocking site $m_b$ in the charge $Q=3^{M-1}$ sector, which is localized on sites $m\le m_b$ and conditionally extended on sites $m>m_b$. (c) In a Krylov subspace of a generic charge sector $Q$, the lattice is divided into alternating localized and conditionally extended intervals, leading to MBL eigenstates.
}
\label{fig-MBL}
\end{figure}

\subsubsection{A Krylov subspace without blocking}

\begin{figure}[tbp]
\begin{center}
\includegraphics[width=3.3in]{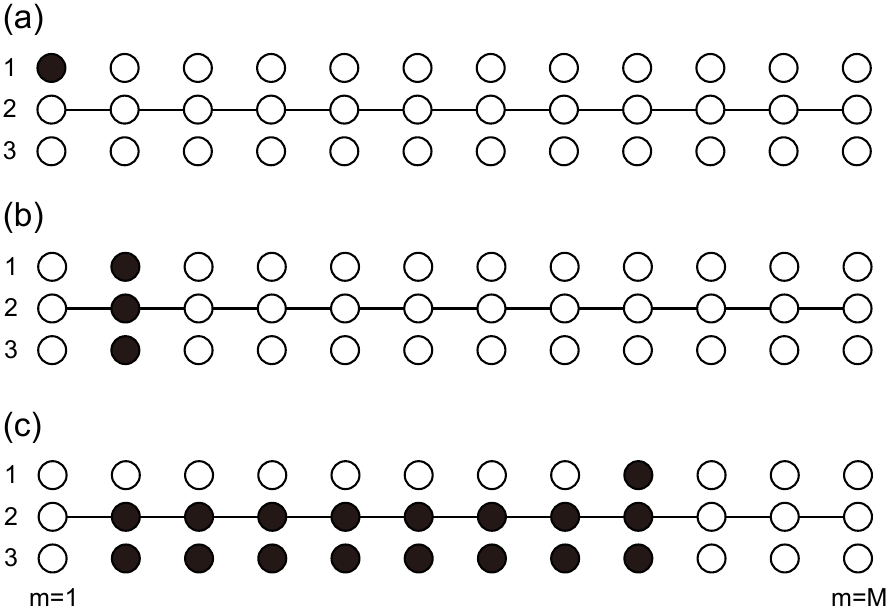}
\end{center}
\caption{Illustration of the Fock states in the Krylov subspace without blocking (as given by \cref{eq:sub-Hil-3-0}) in the charge $Q=3^{M-1}$ sector of the $N=3$ model at $W=0$. The three circles on each site $m$ stand for the fermion basis $f_{m,j}$ after rotation in \cref{eq:H3}. Solid and hollow circles stand for occupied and empty fermion modes, respectively. (a)-(c) shows three possible fermion occupation states in this Krylov subspace.
}
\label{fig-Ne3-krylov1}
\end{figure}

A \emph{Krylov subspace} is defined as a closed Hilbert space spanned by states $H^{n}|\psi\rangle$ of all nonnegative integers $n\ge0$ from some root state $|\psi\rangle$, where $H$ is the Hamiltonian. We first examine a special Krylov subspace in the $Q=3^{M-1}$ charge sector, which has a basis of $M$ Fock states defined by
\begin{equation}\label{eq:sub-Hil-3-0}
|m_s,1,1\rangle=f_{m_s,1}^\dag\prod_{m=2}^{m_s}f^\dag_{m,2}f^\dag_{m,3}|0\rangle\ ,
\end{equation}
where $1\le m_s\le M$. Note that the first fermion mode $f_{m,1}$ on each site $m$ is empty except for the $m_s$-th site. The fermion occupations of such states in \cref{eq:sub-Hil-3-0} are illustrated in \cref{fig-Ne3-krylov1}. The 3-quantum number notation of state $|m_s,1,1\rangle$ will become clear in next subsection. These $M$ Fock states form a closed Hilbert Krylov subspace as illustrated by \cref{fig-Ne3-krylov1}, in which we choose $|1,1,1\rangle$ to be the root state, and the Hamiltonian in Eq. (\ref{eq:H3}) acts in this Krylov subspace as
\begin{equation}\label{eq:H-tb-0}
\begin{split}
H&|m_s,1,1\rangle=\vartheta_{m_s-1}\sqrt{3}J_{m_s-1}|m_s-1,1,1\rangle \\
& + \vartheta_{M-m_s}\sqrt{3}J_{m_s}|m_s+1,1,1\rangle+V_{m_s}|m_s,1,1\rangle.
\end{split}
\end{equation}
Here $V_{m_s}=\mu_{m_s}+2\sum_{m=2}^{m_s}\mu_m$, the coefficients $J_m$ are given in \cref{eq-Jm}, and the notation $\vartheta_m$ is defined in Eq. (\ref{eq-heaviside}). Such a Hamiltonian is thus equivalent to a single-particle tight-binding Hamiltonian on sites $1\le m_s\le M$ with an open boundary condition, which has an on-site potential $V_{m_s}$, and a nearest neighbor hopping  $\sqrt{3}J_{m_s}$. The many-body state $|m_s,1,1\rangle$ is thus equivalent to a single free effective particle on site $m_s$. We say this subspace has no blocking (also defined in the next subsection), since all the sites are connected by hopping.

It is straightforward to diagonalize such a single-particle Hamiltonian and obtain the eigenstates in this subspace. Generically, if the parameters $V_{m_s}$ and/or $J_{m_s}$ are disordered with respect to $m_s$, the tight-binding Hamiltonian in \cref{eq:H-tb-0} will exhibit the \emph{Anderson localization} \cite{anderson1958}, with all the eigenstates spatially localized around different sites $m_s$. Since the basis in \cref{eq:sub-Hil-3-0} are many-body fermion states, these eigenstates should be viewed as many-body localized (MBL) in the original fermion basis.

In the case when \cref{eq-muJ} is satisfied, namely, when the chemical potential $\mu_m=\mu$ and interaction strength $J_m=J$ are uniform, one has $V_m=(2m-1)\mu$ that increases linearly with $m$. If $\mu\neq0$, the Hamiltonian of \cref{eq:H-tb-0} resembles that of the non-interacting electron Bloch oscillation in electric field, and thus the eigenstates will be \emph{Wannier-Stark localized} \cite{wannier1962} within roughly a localization length of order $J/|\mu|$.

In the special case when
\begin{equation}\label{eq:uniform-J-N=3}
\mu_m=\frac{\mu_1}{3^{m-1}}\ ,\qquad J_m=J\ ,
\end{equation}
one has $V_{m_s}=\mu_1$ and $J_{m_s}=J$ independent of $m_s$. As a result, the eigenstates will be extended in the entire 1D chain, given by:
\begin{equation}
|\psi_{k,1,1}\rangle=\sum_{m_s=1}^{M} \sin \frac{\pi k m_s}{M+1} |m_s,1,1\rangle\ ,
\end{equation}
where $k$ is an integer ($1\le k\le M$). The according eigenenergy is 
\begin{equation}\label{eq-N3-Ek11}
E_{k,1,1}=\mu_1+2\sqrt{3}J\cos  \frac{\pi k}{M+1}\ .
\end{equation}
More generally, the eigenstates will be extended if $V_{m}$ and $J_m$ are periodic in $m$.

Therefore, we say the eigenstates in this Krylov subspace of Eq. (\ref{eq:sub-Hil-3-0}) are \emph{conditionally extended} (on the hollow sites in Fig. \ref{fig-MBL}(a)), meaning that they are extended if and only if $V_{m}$ and $J_m$ are periodic in $m$. The states in this Krylov subspace behave as a single free effective particle in the entire lattice.

%It is easy to see that if a many-body Fock state $|\Psi\rangle$ satisfies 

\subsubsection{Other Krylov subspaces: blocking enforced localization}\label{sec:N=3-W=0-ss2}

Following a similar idea, we can determine all the other Krylov subspaces in the charge $Q=3^{M-1}$ sector and derive the eigenstates exactly.

\begin{figure}[tbp]
\begin{center}
\includegraphics[width=3.3in]{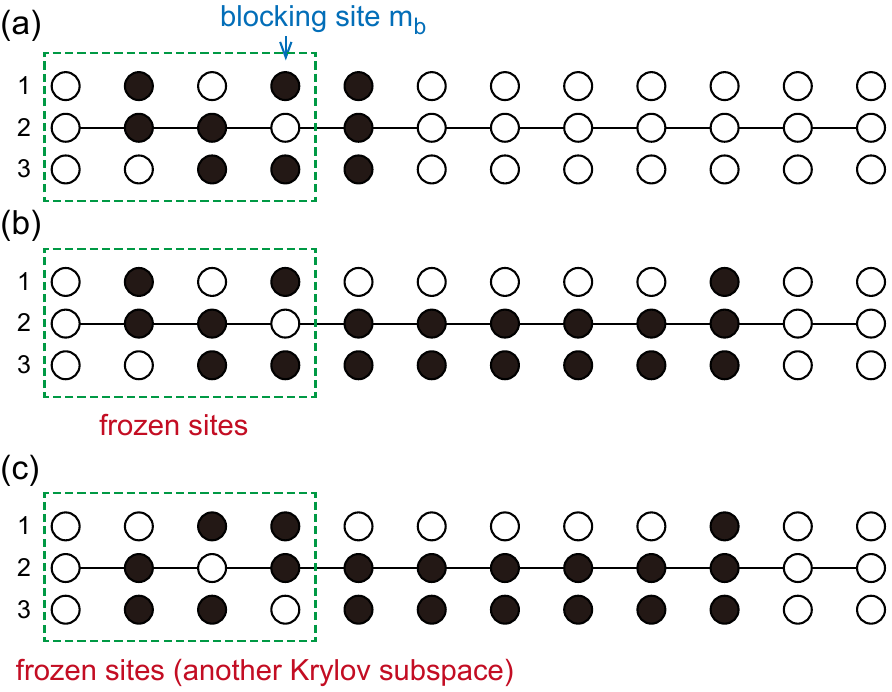}
\end{center}
\caption{Illustration of the Krylov subspace with a blocking site $m_b$ (where the first fermion mode $f_{m_b,1}$ is occupied) in the charge $Q=3^{M-1}$ sector of the $N=3$ model at $W=0$. Solid and hollow circles stand for occupied and empty fermion modes, respectively. The fermion occupation on sites $m\le m_b$ is frozen. (a)-(b) shows two possible fermion occupation states in such a Krylov subspace, with (a) being the root state. (c) shows a state in another Krylov subspace with the same blocking site $m_b$ but a different frozen fermion occupation on sites $m\le m_b$.
}
\label{fig-Ne3-krylov2}
\end{figure}

The Fock basis in \cref{eq:sub-Hil-3-0} in the Krylov subspace above has the first fermion mode $f_{m,1}$ empty in all sites but site $m_s$. We now consider a set of more generic Fock states defined as follows:
\begin{equation}\label{eq:sub-Hil-3-mb}
\begin{split}
&|m_s,m_b,\{s_m^{j}\}\rangle \\
=&f_{m_s,1}^\dag\prod_{m=m_b+1}^{m_s}f^\dag_{m,2}f^\dag_{m,3}\prod_{m=2}^{m_b}f^\dag_{m,s^1_m}f^\dag_{m,s^2_m}|0\rangle,
\end{split}
\end{equation}
where $2\le m_b<m_s\le M$, and $\{s_m^{j}\}$ are the indices of occupied fermion modes on sites $2\le m\le m_b$ ($j=1,2$), with the specific requirement on site $m_b$:
\begin{equation}
s_{m_b}^1\equiv1\ , \qquad s_{m_b}^2=2\text{ or }3\ .
\end{equation}
For all the other sites $2\le m< m_b$, the coefficients $s_m^1$ and $s_m^2$ can take any two distinct values among $\{1,2,3\}$. Therefore, $m_b$ and $m_s$ are the last two sites with the first fermion mode $f_{m,1}$ occupied. Some examples of Fock states in \cref{eq:sub-Hil-3-mb} are given in \cref{fig-Ne3-krylov2}.

We call the site $m_b$ in \cref{eq:sub-Hil-3-mb} a \emph{blocking site}, for the reasons below. Since the first fermion mode $f_{m_b,1}$ on site $m_b$ is occupied, and the $(m_b+1)$-th site is not empty, it is easy to see that the Hamiltonian in \cref{eq:H3} cannot change the occupation configuration on sites $1\le m\le m_b$ by any number of actions on the Fock state in \cref{eq:sub-Hil-3-mb}. Namely, the coefficients $\{s_m^{j}\}$ are invariant, and the fermions on sites $1\le m\le m_b$ are frozen, as illustrated in \cref{fig-Ne3-krylov2}. Therefore, the dynamics of the Fock state in \cref{eq:sub-Hil-3-mb} is blocked at site $m_b$ and only happens on sites $m>m_b$.

As a result, one can show that all the Fock states $|m_s,m_b,\{s_m^{j}\}\rangle$ in \cref{eq:sub-Hil-3-mb} with the same blocking site $m_b$ and occupation configuration $\{s^j_m\}$ form a closed Krylov subspace, and we define the root state of this Krylov subspace to be the Fock state with $m_s=m_b+1$ (\cref{fig-Ne3-krylov2}(a)). So we can simply label the Krylov subspaces by quantum numbers $(m_b,\{s_m^{j}\})$. The Hamiltonian in \cref{eq:H3} acting on Fock states in such a Krylov subspace in a similar way to \cref{eq:H-tb-0}, but only on sites $m_s>m_b$:
\begin{equation}\label{eq:H-tb-mb}
\begin{split}
H & |m_s, m_b,\{s_m^{j}\}\rangle=V_{m_s} |m_s,m_b,\{s_m^{j}\}\rangle\\
&\quad+\vartheta_{m_s-m_b-1}\sqrt{3}J_{m_s-1}|m_s-1,m_b,\{s_m^{j}\}\rangle \\
&\quad+\vartheta_{M-m_s} \sqrt{3}J_{m_s}|m_s+1,m_b,\{s_m^{j}\}\rangle\ ,
\end{split}
\end{equation}
where as before, $V_{m_s}=\mu_{m_s}+2\sum_{1<m\le m_s}\mu_m$, $J_m$ is defined in \cref{eq-Jm}, and $\vartheta_m$ is given in \cref{eq-heaviside}. Therefore, in the Krylov subspace of Fock states in \cref{eq:sub-Hil-3-mb} with fixed indices $m_b$ and $\{s_m^{j}\}$, the Hamiltonian behaves as a single-particle tight-binding model on sites $m_b+1\le m_s\le M$ with an open boundary condition, as shown in \cref{fig-Ne3-krylov2}(a)-(b), and the states can be viewed as a single free effective particle in this interval. In particular, the energy spectrum of \cref{eq:H-tb-mb} for a fixed blocking site $m_b$ is obviously independent of the indices $\{s_m^{j}\}$. Therefore, each energy level with a blocking site $m_b$ in the charge $Q=3^{M-1}$ sector has a degeneracy $2\cdot 3^{m_b-2}$, which is the number of possible frozen index configurations $\{s_m^{j}\}$, or the number of Krylov subspaces with the same blocking site $m_b$.

The eigenstates of \cref{eq:H-tb-mb} in a Krylov subspace with blocking site $m_b$ are thus conditionally extended on sites $m_b<m\le M$, as illustrated in Fig. \ref{fig-MBL}(b) and \cref{fig-Ne3-krylov2}. If $V_{m_s}$ and $J_{m_s}$ are not periodic in $m_s$, one expects the eigenstates to be Anderson localized (or Wannier-Stark localized if $V_{m_s}$ is linear in $m_s$), or MBL in the fermion basis perspective. If instead $V_{m_s}$ and $J_{m_s}$ are periodic, for instance if \cref{eq:uniform-J-N=3} is satisfied, yielding $V_{m_s}=\mu_1$ and $J_{m_s}=J$, the eigenstates will be extended within $m_b<m\le M$, with wavefunctions
\begin{equation}
|\psi_{k,m_b,\{s^j_m\}}\rangle =\sum_{m_s=m_b+1}^{M} \sin \frac{\pi k (m_s-m_b)}{M-m_b+1} |m_s,m_b,\{s_m^{j}\}\rangle,
\end{equation}
where $k$ is an integer ranging within $1\le k\le M-m_b$. The energy of this eigenstate is given by 
\begin{equation}\label{eq-N3-Ekmbs}
E_{k,m_b,\{s^j_m\}}=\mu_1+2\sqrt{3}J\cos  \frac{\pi k}{M-m_b+1}\ , 
\end{equation}
which is degenerate with respect to all the $2\cdot 3^{m_b-2}$ frozen occupation configurations $\{s^j_m\}$, as we have explained.

In addition to the above eigenstates in Krylov spaces with blocking site $m_b\ge 2$, there are two Fock states decoupled from all the other states:
\begin{equation}\label{eq:sub-Hil-3-1}
|1,1,s\rangle=f^\dag_{1,s}|0\rangle\ ,\quad (s=2,3)\ ,
\end{equation}
which are readily the eigenstates of the Hamiltonian in \cref{eq:H3}. Their energy eigenvalues are simply
\begin{equation}\label{eq-N3-Ek1s}
E_{1,1,s}=\mu_1\ ,\qquad (s=2,3).
\end{equation}
We can view these two states as two one-dimensional Krylov subspaces with a blocking site $m_b=1$. In this way, we can summarize the Fock states in \cref{eq:sub-Hil-3-0,eq:sub-Hil-3-mb,eq:sub-Hil-3-1} in the same notation $|m_s,m_b,\{s_m^{j}\}\rangle$, with $1\le m_b\le m_s\le M$, and $(m_b,\{s_m^{j}\})$ label the Krylov subspace. If $m_b\ge 2$, only those states with $m_b<m_s$ are valid.

One can verify that the total number of eigenstates we have found adds up to the Hilbert space dimension $h_{Q}=\frac{3^{M-1}+5}{2}$ of the charge sector $Q=3^{M-1}$.

\subsubsection{Numerical spectrum and entanglement entropy}\label{sec:N=3-W=0-numer}

To verify the above exact solutions in the $Q=3^{M-1}$ sector, we perform exact diagonalization (ED) calculations of the energy spectrum of the $N=3$ quantum breakdown model. For generality, our ED is performed in a full charge $Q$ sector in the original fermion basis $c_{m,i}$, assuming not knowing any Krylov subspaces.

\begin{figure}[tbp]
\begin{center}
\includegraphics[width=3.3in]{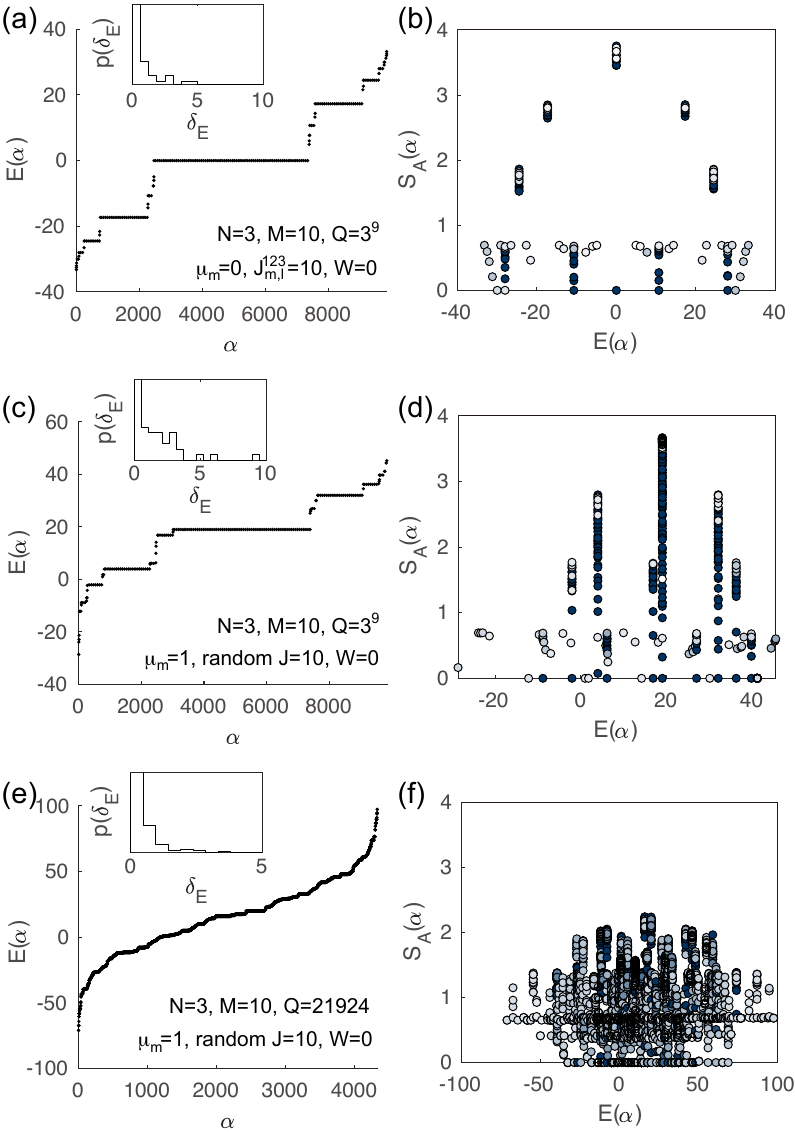}
\end{center}
\caption{ED calculation in a charge $Q$ sector of the model with $(N,M)=(3,10)$ and disorder potential $W=0$. (a), (c), (e) shows the energy levels $E(\alpha)$ sorted from low to high (labeled by $\alpha$), and the corresponding LSS in the insets. (b), (d), (f) show the eigenstate entanglement entropy in subregion $A$ (the first $\lfloor M/2\rfloor$ sites) versus energy $E(\alpha)$ for (a), (c), (e), respectively (the darker the higher density of dots). The parameters are labeled in the left panels. In (a)-(b), the interaction $J^{123}_{m,l}=10$ is uniform, and $Q=3^{M-1}$. In (c)-(f), the interaction $J^{123}_{m,l}$ is random with strength $J=10$, while $Q=3^{M-1}$ in (c)-(d) and $Q=21924$ in (e)-(f). 
}
\label{figN=3-W=0}
\end{figure}

We first show an example with $(N,M)=(3,10)$ and no randomness: we assume all the interactions in \cref{eq-HI} are uniformly $J^{123}_{m,l}=10$, the chemical potential $\mu_m=\mu=0$, and the disorder potential $W=0$. The many-body energy spectrum in the $Q=3^{M-1}=3^9$ sector is shown in Fig. \ref{figN=3-W=0}(a), where $\alpha$ is an integer labeling the energy levels $E(\alpha)$ sorted from low to high ($1\le \alpha\le h_Q$). The energy levels agree well with \cref{eq-N3-Ek11,eq-N3-Ekmbs,eq-N3-Ek1s}. Not surprisingly, most energy levels have an extensive degeneracy, making the spectrum fractal-like. This is consistent with the fact that the energy levels with a blocking site $m_b$ have a degeneracy $2\cdot 3^{m_b-2}$ (the number of degenerate Krylov subspaces with the same $m_b$) plus other accidental degeneracies, and the fractal-like spectrum is because the degeneracies are proportional to powers of $3$. For instance, the degeneracy of the $E=0$ mode in Fig. \ref{figN=3-W=0}(a) is contributed by all the states of certain $m_b$ with $k=(M-m_b+1)/2\in\mathbb{Z}$ in \cref{eq-N3-Ekmbs}. Moreover, we note that the parameters of Fig. \ref{figN=3-W=0}(a) satisfy \cref{eq:uniform-J-N=3} with $\mu_1=0$ and $J=10$, so the eigenstates are extended on sites $m>m_b$.

Another example we show in the $(N,M)=(3,10)$ case has random complex interactions $J^{ijk}_{m,l}$ satisfying \cref{eq-J}, where the interaction strength $J=10$, while the chemical potential $\mu_m=\mu=1$ is uniform, and $W=0$. Fig. \ref{figN=3-W=0}(c) shows the energy spectrum in the $Q=3^{M-1}=3^9$ sector. Clearly, the fractal-like level degeneracies remain for random interactions, agreeing well with the degeneracy $2\cdot 3^{m_b-2}$ for blocking site $m_b$. The randomness of interaction does remove all accidental degeneracies.

The level spacing statistics (LSS) in a charge sector is known as an indicator of integrability or quantum chaos, which is defined as the statistical distribution $p(\delta_E)$ of the nearest neighboring level spacing 
\begin{equation}\label{eq:spacing}
\delta_E(\alpha)=E(\alpha+1)-E(\alpha)\ ,\quad (1\le \alpha\le h_Q-1)\ .
\end{equation}
Integrable systems show Poisson distribution LSS \cite{berry1977} 
\begin{equation}
p(\delta_E)\propto e^{-\delta_E/\lambda_0}\ ,
\end{equation}
with some constant $\lambda_0$, while fully quantum chaotic systems show Wigner-Dyson (WD) distribution LLS \cite{bohigas1984,Wigner1967,Dyson1970}
\begin{equation}
p(\delta_E)\propto \delta_E^{n} e^{-\delta_E^2/\lambda_0^2} \ ,
\end{equation}
where $n=1,2,4$ for models in the Gaussian orthogonal ensemble (GOE), Gaussian unitary ensemble (GUE) and Gaussian symplectic ensemble (GSE) symmetry classes, respectively.

The insets of Fig. \ref{figN=3-W=0}(a) and (c) show the LSS in the charge $Q=3^9$ sector. The extensive level degeneracies yield a delta function peak in the LSS at $\delta_E=0$, while at $\delta_E>0$, the LSS shows a Poisson distribution (although $\delta_E$ is large compared to $1/h_Q$). This is consistent with the fact that all the eigenstates in the $Q=3^{M-1}$ sector are exactly solvable, namely, integrable.

To tell how localized the eigenstates are, we calculate the entanglement entropy 
\begin{equation}
S_A(\alpha)=-\text{tr}[\rho_A(\alpha)\ln\rho_A(\alpha)]
\end{equation}
of the reduced density matrix $\rho_A(\alpha)$ of each eigenstate $|\alpha\rangle$ in a subregion $A$. The subregion $A$ is always defined as sites $1\le m\le\lfloor M/2\rfloor$ hereafter, where $\lfloor x\rfloor$ is the floor function giving the integer part of $x$. MBL and extended eigenstates should exhibit area law and volume law entanglement entropies, respectively. Fig. \ref{figN=3-W=0}(b) and (d) show the entanglement entropy $S_A(\alpha)$ of the $Q=3^{M-1}$ sector eigenstates in Fig. \ref{figN=3-W=0}(a) and (c), respectively. Each dot represents an eigenstate, and the darker the color of the dots is, the denser the dots are (to reflect the overlapping dots). In both cases, there are a large number of eigenstates with low entanglement entropy irrespective of the eigen-energy $E(\alpha)$, which can be identified as area-law MBL eigenstates. The rest states, however, have higher entanglement entropy peaked in the middle of the energy spectrum, showing a volume law. The presence of volume law states has two reasons: first, as shown in Fig. \ref{fig-MBL}, many eigenstates have conditionally extended regions, which may show a relatively large entanglement. In particular, compared to most other charge sectors, the $Q=3^{M-1}$ has a relatively larger conditionally extended region (see Sec. \ref{sec:N=3-generic-Q} below), thus is more delocalized. In Fig. \ref{figN=3-W=0}(b) where the extended condition \cref{eq:uniform-J-N=3} is satisfied, the eigenstates are like an extended free effective particle in the conditionally extended region. In Fig. \ref{figN=3-W=0}(d) where the chemical potential $\mu_m=\mu=1$, one expects Wannier-Stark localization with localization length $\sim J/|\mu|$, thus entanglement entropies are generically lower than those in Fig. \ref{figN=3-W=0}(b). Second, although the eigenstates in each Krylov subspace we give in Sec. \ref{sec:N=3-W=0-ss2} are strictly localized on sites $m<m_b$, one can linearly superpose the extensively degenerate eigenstates from different Krylov subspaces into a new eigenbasis with high entanglement entropies. This is generically the case in ED which ends up with a random degenerate eigenbasis. Intriguingly, adding a small disorder potential $W$ (Eq. \ref{eq-W}) is sufficient to slightly break the degeneracy and fix the eigenbasis to a low-entanglement basis, as shown in Fig. \ref{figN=3-W>0}(b), which has $W=10^{-6}$ and the rest parameters the same as those in Fig. \ref{figN=3-W=0}(b). Moreover, if one rotates the fermion basis to the $f_{m,l}$ basis in \cref{eq:H3}, adding a small disorder potential diagonal in the $f_{m,l}$ basis pins all the eigenstates into each Krylov subspace we described, which have entanglement entropy no larger than $\ln 2$ (see \cref{app:rot-basis-N=3}) for details).

\subsection{The generic charge sectors}\label{sec:N=3-generic-Q}

We now study the Krylov subspaces and eigenstates in an arbitrary charge $Q$ sector of the $N=3$ model. Similar to \cref{eq-config-N=3}, the allowed fermion number configurations $\{n_m\}$ of the on-site Fock basis are uniquely determined by charge $Q$. For later purpose, we rewrite the charge $Q$ as
\begin{equation}\label{eq:Qgeneral} 
Q=\sum_{m=1}^{M} n^0_m 3^{M-m}\ ,
\end{equation}
where the integers $n^0_m$ satisfy the following condition:
\begin{equation}\label{eq:Qgeneral-nm0} 
n^0_m\begin{cases}
&=3 \quad (m\le m_c)\ , \\
&\in \{0,1,2\}\ ,\quad (m>m_c)\ ,\\
\end{cases}
\end{equation}
and $m_c\ge0$ is an integer uniquely determined by $Q$. We call $\{n^0_m\}$ the \emph{reference configuration} of charge $Q$, which has one-to-one unique correspondence with charge $Q$. For instance, the $Q=3^{M-1}$ sector we studied in Sec. \ref{sec:sectorQ1-N=3} corresponds to $\{n^0_m\}=\{1,0,0,\cdots\}$. Intuitively, the reference configuration $\{n^0_m\}$ of charge $Q$ in \cref{eq:Qgeneral-nm0} is the configuration with the smallest possible total number of fermions in the charge $Q$ sector. In other words, $\{n^0_m\}$ in \cref{eq:Qgeneral-nm0} is the charge configuration where no fermions can move any further to the left via interaction.

In particular, for any charge $Q<3^{M}$, one has $m_c=0$, and thus $0\le n^0_m\le 2$ for all $m$. In this case, the set of integers $\{n^0_m\}$ is nothing but the digits of $Q$ in ternary number. A few examples of the reference configurations $\{n_m^0\}$ of charge $Q$ sectors in the $(N,M)=(3,10)$ case are shown in \cref{Tab-Q}.

\begin{table}[tbp]
\centering
\begin{tabular}{c|c|c|c}
\hline
number of sites $M$ & charge $Q$ & $\{n_m^0\}$ & $m_c$  \\
\hline
$10$ & $3^9$ & $1,0,0,0,0,0,0,0,0,0$ & $0$  \\
\hline
$10$ & $21924$ & $1,0,1,0,0,0,2,0,0,0$ & $0$  \\
\hline
$10$ & $61813$ & $3,0,1,0,2,1,0,1,0,1$ & $1$  \\
\hline
$10$ & $86241$ & $3,3,3,1,0,2,2,0,1,0$ & $3$  \\
\hline
\end{tabular}
\caption{ Examples of reference configurations $\{n_m^0\}$ of charge $Q$ sectors for $(N,M)=(3,10)$. The corresponding $m_c$ is also given according to \cref{eq:Qgeneral-nm0}.}\label{Tab-Q}
\end{table}

%Similar to \cref{eq-config-N=3}, the allowed fermion number configurations $\{n_m\}$ of the on-site Fock basis are uniquely determined by charge $Q$. In particular, the reference configuration $\{n^0_m\}$ of charge $Q$ in \cref{eq:Qgeneral} is the configuration with the smallest total number of fermions possible for charge $Q$.

We first note that if $m_c>1$ in \cref{eq:Qgeneral-nm0}, the fermions on the fully filled sites $m<m_c$ of the reference configuration are frozen, so we can neglect them in the analysis of Krylov subspaces. Then, we assume the reference configuration has $M_Q$ partially filled sites on which $0<n^0_m<3$, and we assume $m_{f,p}$ is the $p$-th such partially filled site sorted from small to large ($1\le p\le M_Q$). We now show that depending on the occupation numbers $n^0_{m_{f,p}}$ on these $M_Q$ sites, the eigenstates in the charge $Q$ sector fractionalizes into an exponentially large number of Krylov subspaces which are either exactly solvable or almost solvable.

\subsubsection{Fully exactly solvable charge sectors}\label{sec:N=3-generic-Q-exact}

%$m^{\text{nz}}_{p}$

Assume charge $Q$ in \cref{eq:Qgeneral} has $n^0_{m_{f,p}}=2$ for any partially filled site $2\le p\le M_Q$ if $m_c=0$, or for any $1\le p\le M_Q$ if $m_c\ge1$. In this case, the eigenstates of all the Krylov subspaces in the charge $Q$ sector can be exactly solved.

To see this, we first note that the $M_Q$ sites at positions $m_{f,p}$ separate the unfrozen part of the system $m_c\le m\le M$ into $M_Q+1$ decoupled subsystems. The first subsystem is the interval of sites $m_c\le m\le m_{f,1}-1$ (which vanishes if $m_{f,1}=1$), and stays trivially in the zero fermion state. The $(p+1)$-th ($1\le p\le M_Q$) subsystem is the interval of sites $m\in [m_{f,p}, m_{f,p+1}-1]$ (here we define $m_{f,M_Q+1}=M+1$). It is easy to verify that the allowed Fock basis fermion number configurations within the $(p+1)$-th subsystem ($1\le p\le M_Q$) are similar to \cref{eq-config-N=3}, given by
\begin{equation}\label{eq-config-N=3-generalQ-exact}
%\begin{split}
n_m=2\vartheta_{m_{s,p}-m+1}-(3-n^0_{m_{f,p}})\delta_{m,m_{f,p}}+\delta_{m,m_{s,p}}
%\end{split}
\end{equation}
for sites $m\in [m_{f,p}, m_{f,p+1}-1]$, where the integer $m_{s,p}\in[m_{f,p}, m_{f,p+1}-1]$ is a free parameter, and $\vartheta_m$ is defined in \cref{eq-heaviside}. Namely, the allowed configurations within the $(p+1)$-th subsystem typically take the form $\{n_m\}=\{n^0_{m_{f,p}}-1,2,\cdots,2,3,0,\cdots, 0 \}$, where the site with $3$ fermions is the  $m_{s,p}$-th site. Note that the allowed charge configuration in the $(p+1)$-th subsystem is solely determined by $n^0_{m_{f,p}}$. This is because $n^0_{m_{f,p}}=2$ for any $p\ge1$ guarantees that the $m_{f,p}$-th site is never empty, so the fermions in the $p$-th subsystem is blocked and cannot spread into with the $(p+1)$-th subsystem. 

\begin{figure}[tbp]
\begin{center}
\includegraphics[width=3.3in]{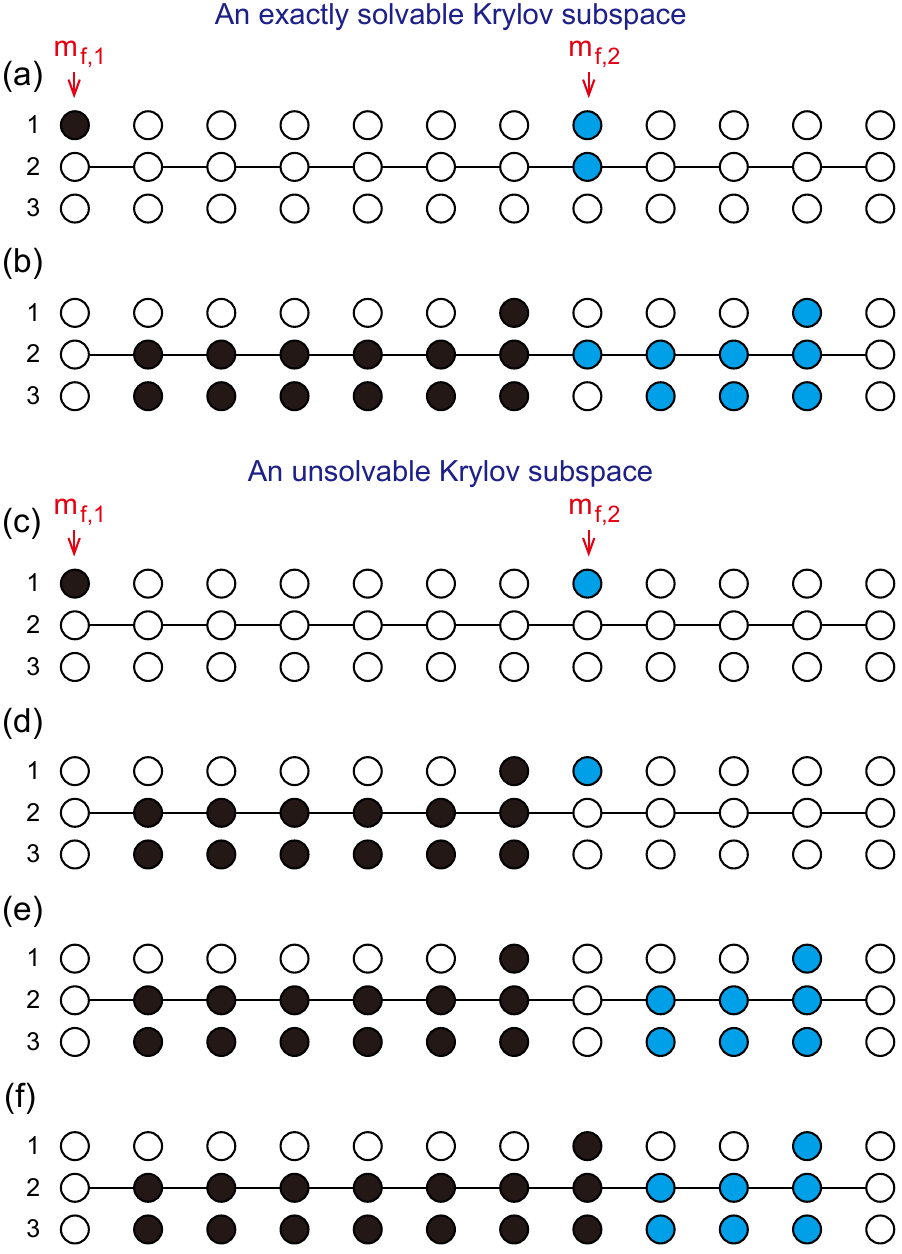}
\end{center}
\caption{Illustration of Krylov subspaces without blocking sites in generic charge $Q$ sectors with two filled sites $m_{f,1}$ and $m_{f,2}$ in the reference configuration, for the $N=3$ model at $W=0$. Solid and hollow circles stand for occupied and empty fermion modes, respectively. (a) shows the reference configuration of a charge $Q$ sector with 2 fermions on site $m_{f,2}$, and (b) shows a generic state in this Krylov subspace. In this case, site $m_{f,2}$ divides the system into two decoupled subsystems which are exactly solvable. (c) shows the reference configuration of another charge $Q$ sector with 1 fermion on site $m_{f,2}$. (d)-(f) show three possible states in this Krylov subspace. In this case, site $m_{f,2}$ can be occupied by either fermions from the right (blue) and from the left (black), yielding an effective hard core interaction between the left and right subsystems.
}
\label{fig-Ne3-krylov3}
\end{figure}

An explicit example is given in \cref{fig-Ne3-krylov3}(a)-(b), where the reference configuration is shown in \cref{fig-Ne3-krylov3}(a), which has $m_c=0$ and two partially occupied sites $m_{f,1}$ and $m_{f,2}$, with $1$ (black) and $2$ (blue) fermions, respectively. As shown in \cref{fig-Ne3-krylov3}(b), under interactions, the site $m_{f,2}$ always has at least one fermion occupied, thus blocking any fermions from sites $m<m_{f,2}$ to spread into or beyond site $m_{f,2}$. The entire system is thus divided into two decoupled subsystems on sites $m_{f,1}\le m<m_{f,2}$ and $m_{f,2}\le m<M$, respectively.

Because of the decoupling of the $M_Q+1$ subsystems, the many-body eigenstates of the entire system decomposes into the direct products of $M_Q+1$ subsystem eigenstates. 
%Except for the first subsystem which is trivially in the particle vacuum state, 
Each subsystem has Krylov subspaces and exactly solvable eigenstates by the same method as we used for the charge $Q=3^{M-1}$ sector (see Sec. \ref{sec:sectorQ1-N=3}). More explicitly, a Krylov subspace of the $(p+1)$-th subsystem is labeled by a blocking site $m_{b,p}$, defined as the second last site in the interval $[m_{f,p},m_{f,p+1}-1]$ with the first fermion mode $f_{m_{b,p},1}$ occupied (similar to \cref{eq:sub-Hil-3-mb}), and frozen occupation indices $\{s_m^j\}$ in the interval $[m_{f,p},m_{b,p}]$. Its root state is defined as the Fock state with the smallest possible total fermion number in this subspace. In this Krylov subspace, the Hamiltonian maps to a single-particle tight-binding Hamiltonian similar to \cref{eq:H-tb-0,eq:H-tb-mb} with $m_s$ and $m_b$ replaced by $m_{s,p}$ and $m_{b,p}$ (see \cref{app:any-charge-sector-N=3}), thus is exactly solvable. The Krylov subspaces of the entire system are simply the direct product of the Krylov subspaces of the $M_Q+1$ subsystems.

Fig. \ref{fig-MBL}(c) illustrates the localization of a generic many-body eigenstate in a Krylov subspace of such exactly solvable charge $Q$ sectors. Within the $(p+1)$-th subsystem on sites $[m_{f,p}, m_{f,p+1}-1]$, there is a blocking site $m_{b,p}$. The subsystem is then frozen in the localized configuration $\{s_m^j\}$ within the interval of sites $[m_{f,p},m_{b,p}]$, and maps to a conditionally extended single free effective particle within sites $[m_{b,p}+1,m_{f,p+1}-1]$, similar to the charge $Q=3^{M-1}$ sector (Fig. \ref{fig-MBL}(b)). This separates the entire system into alternating frozen (thus localized) and conditionally extended intervals (Fig. \ref{fig-MBL}(c)), which maps to $M_Q$ free effective particles (one in each conditionally extended interval) decoupled with each other. Such many-body eigenstates are therefore necessarily MBL in the original fermion basis, with a localization length not exceeding the size of each subsystem. In the $Q=3^{M-1}$ sector, there is only one subsystem and some eigenstates can still be extended; while in most charge $Q$ sectors, the number of subsystems $M_Q$ is usually a large number of the same order as $M$, so the localization length is at most $M/M_Q$ which is small.

%We now define $m_{f,p}$ as the $p$-th site (sorted from small to large) that has nonzero parameter $n^0_{m^{\text{nz}}_{p}}>0$, and we assume there are $M_Q$ such sites (thus $1\le p\le M_Q$). Starting from the configuration $\{n^0_m\}$, if $m^{\text{nz}}_{p+1}>m^{\text{nz}}_{p}+1$ (for any $p$), one can replace the fermion number on the sites from $m^{\text{nz}}_{p}$ to $m^{\text{nz}}_{p+1}-1$, i.e., the number string of the form $\{n^0_{m^{\text{nz}}_{p}},0,0,\cdots, 0 \}$, by any number string of the form $\{n^0_{m^{\text{nz}}_{p}}-1,2,\cdots,2,3,0,\cdots, 0 \}$, to arrive at a new allowed fermion number configuration. %This allows one to write down all the allowed fermion number configurations.

%Note that such allowed fermion number strings on the sites from $m^{\text{nz}}_{p}$ to $m^{\text{nz}}_{p+1}-1$ are of the same form as that defined in \cref{eq-config-N=3}, except that here they are defined on a subset of $m^{\text{nz}}_{p+1}-m^{\text{nz}}_{p}$ sites. Therefore, for a generic charge sector $Q$, it is tempting to divide the entire chain of $M$ sites into $M_Q$ decoupled subsystems, the $p$-th of which ranges from sites $m^{\text{nz}}_{p}$ to $m^{\text{nz}}_{p+1}-1$, after which one may solve the eigenstates in each subsystem by applying the method of reduction into effective single-particle tight-binding models as shown in \cref{sec:sectorQ1-N=3}. The eigenstates of the total system would then be the direct product of the eigenstates of all the $M_Q$ subsystems.

\subsubsection{The other charge sectors: exactly solvable in most Krylov subspaces}\label{sec:N=3-generic-Q-approx}

For a generic charge $Q$ sector, where some partially filled sites $m>m_c$ in \cref{eq:Qgeneral-nm0} have fermion number $n_m^{0}=1$, the above picture of $M_Q+1$ decoupled subsystems separated by sites $m_{f,p}$ is no longer exact for all Krylov subspaces, but is still a good approximation. 

A Krylov subspace is still labeled by the blocking sites $\{m_{b,p}\}$ and occupation indices $\{s_m^j\}$ in the $M_Q$ intervals $[m_{f,p},m_{b,p}]$. We still define the $(p+1)$-th subsystem as the interval $[m_{f,p},m_{f,p+1}-1]$. If one has $n_{m_{f,p}}^0=1$ on a site $m_{f,p}$ for some $p\ge 2$ (or $p=1$ if $m_c\ge1$), there will be a weak interaction between the $p$-th subsystem and the $(p+1)$-th subsystem on their boundary site $m_{f,p}$ in certain Krylov subspaces. This is because $n_{m_{f,p}}^0=1$ allows site $m_{f,p}$ to be either occupied or empty (by the actions of interaction $H_I$ in \cref{eq:H3}). When site $m_{f,p}$ is empty, it can either gain $3$ fermions from the $p$-th subsystem on its left, or gain $1$ fermion from the $(p+1)$-th subsystem on its right, which induces an interaction between the two subsystems on site $m_{f,p}$. Note that the fermions in each of the two subsystems cannot invade into each other beyond the boundary site $m_{f,p}$, which restricts the interaction to be on the site $m_{f,p}$ only. However, such an interaction is present only in the Krylov subspaces with the blocking site $m_{b,p}=m_{f,p}$, namely, with no blocking in the $(p+1)$-th subsystem. This is because if $m_{b,p}>m_{f,p}$, the occupation configuration in the interval $[m_{f,p}+1,m_{b,p}]$ will be fixed, and the site $m_{f,p}$ can only either be empty or accept 3 fermions from the $p$-th subsystem on its left, which eliminates the interaction. Therefore, there are only very few Krylov subspaces having this inter-subsystem boundary interaction, which must have $m_{b,p}=m_{f,p}$. A more detailed expression of such an inter-subsystem interaction can be found in \cref{app:any-charge-sector-N=3}.

\cref{fig-Ne3-krylov3}(c)-(f) shows an example of Krylov subspace of interacting subsystems we described above. \cref{fig-Ne3-krylov3}(c) gives the reference configuration which has two partially filled sites $m_{f,1}$ and $m_{f,2}$ with $1$ fermion each (colored black and blue), respectively.

If one view the Fock states in a Krylov subspace within each interval $[m_{b,p},m_{f,p+1}]$ as a conditionally extended single effective particle on site $m_{s,p}\in[m_{b,p},m_{f,p+1}]$ (the site with $3$ fermions) similar to that in \cref{eq:H-tb-0,eq:H-tb-mb}, the above picture can be illustrated as follows. If $n_{m_{f,p}}^0=1$ and $m_{b,p}=m_{f,p}$ for some $p\ge2$, the $(p-1)$-th effective particle and the $p$-th effective particle will have a hard core repulsion on site $m_{f,p}$. Otherwise, the $(p-1)$-th effective particle and the $p$-th effective particle are decoupled. Namely, among the $M_Q$ effective particles (one in each conditionally extended interval in \cref{fig-MBL}(c)), two neighboring effective particles can have a boundary hard core interaction with each other only when their conditionally extended intervals touch each other and the boundary site $m_{f,p}$ between them has $n_{m_{f,p}}^0=1$. Such an interaction obstructs us to analytically obtain the eigenstates in these Krylov subspaces.

%Such an interaction remains an interacting term in the mapping similar to \cref{eq:H-tb-0,eq:H-tb-mb}. However, this interaction is very weak. This is because, whenever the $p+1$-th subsystem has a blocking site $m_{b,p}$ in the interval $[m_{f,p}+1,m_{f,p+1}-1]$, the site $m_{f,p}$ would only be able to be either empty or to accept 3 fermions from the $p$-th subsystem on its left. Namely, the existence of a blocking site $m_{b,p}>m_{f,p}$ in the $(p+1)$-th subsystem will eliminate the interaction between the $p$-th and the $(p+1)$-th subsystems. Therefore, the interaction is only nonzero when the $(p+1)$-th subsystem is in its subspace without a blocking site, which is a small Hilbert space of the $(p+1)$-th subsystem with dimension at most $m_{f,p+1}-m_{f,p}+1$ (similar to the subspace in \cref{eq:sub-Hil-3-0}). It does interact with the entire Hilbert space of the $p$-th subsystem, so the subspaces with different blocking sites $m_{b,p-1}$ in the $p$-th subsystem are no longer exactly decoupled.

Since this interaction in any generic charge $Q$ sector only weakly exists on the boundary between some neighboring subsystems in very few Krylov subspaces, we expect the eigenstates to be well approximated by direct products of eigenstates of $M_Q+1$ subsystems, which would still show MBL. As a numerical test, in Fig. \ref{figN=3-W=0}(e)-(f) we show the energy spectrum and subregion entanglement entropy of the charge $Q=21924$ sector eigenstates of the $(N,M)=(3,10)$ model calculated by ED, where the model parameters are the same as Fig. \ref{figN=3-W=0}(c)-(d), namely, uniform $\mu_m=\mu=1$, random interaction strength $J=10$ (the same random $J^{ijk}_{m,l}$), and $W=0$. This charge $Q$ in the form of \cref{eq:Qgeneral} corresponds to $\{n^{0}_{1},n^{0}_{3},n^{0}_{7}\}=\{1,1,2\}$ and all the other $n^0_m=0$ (see \cref{Tab-Q}). Fig. \ref{figN=3-W=0}(e) shows that there are still extensive degenerate energy levels. The LSS shows a delta function peak at $\delta_E=0$ and a Poisson distribution at $\delta_E>0$, in consistency with MBL. In Fig. \ref{figN=3-W=0}(f), we see an extensive number of low entanglement entropy eigenstates within the entire energy range, indicating their MBL. The rest states have higher entanglement entropy due to the energy level degeneracies, for the same reason as described in the end of Sec. \ref{sec:N=3-W=0-numer}. But the eigenstate entanglement entropies of the $Q=21924$ sector are generically lower than those in the $Q=3^9$ sector (in Fig. \ref{figN=3-W=0}(d)), since the eigenstates are more localized with multiple nearly decoupled subsystems. Similarly, by rotating to the $f_{m,l}$ basis in \cref{eq:H3}, one can eliminate such degeneracy induced high entanglement entropy in the ED calculation (see \cref{app:rot-basis-N=3} for details).

As our Krylov subspace exact or almost exact solutions and ED calculations show, the $N=3$ quantum breakdown model with $W=0$ shows a 1D MBL in generic charge $Q$ sectors when $M\rightarrow\infty$ (illustrated in \cref{fig-MBL}). Note that such an MBL can exist without any disorder and randomness: we can set $\mu_m=\mu$, $M=0$, and $J^{123}_{m,l}=J$ all being constant and still have MBL.

\section{The $N=3$ model with $W>0$}\label{sec-N=3-W>0}

\begin{figure}[tbp]
\begin{center}
\includegraphics[width=3.3in]{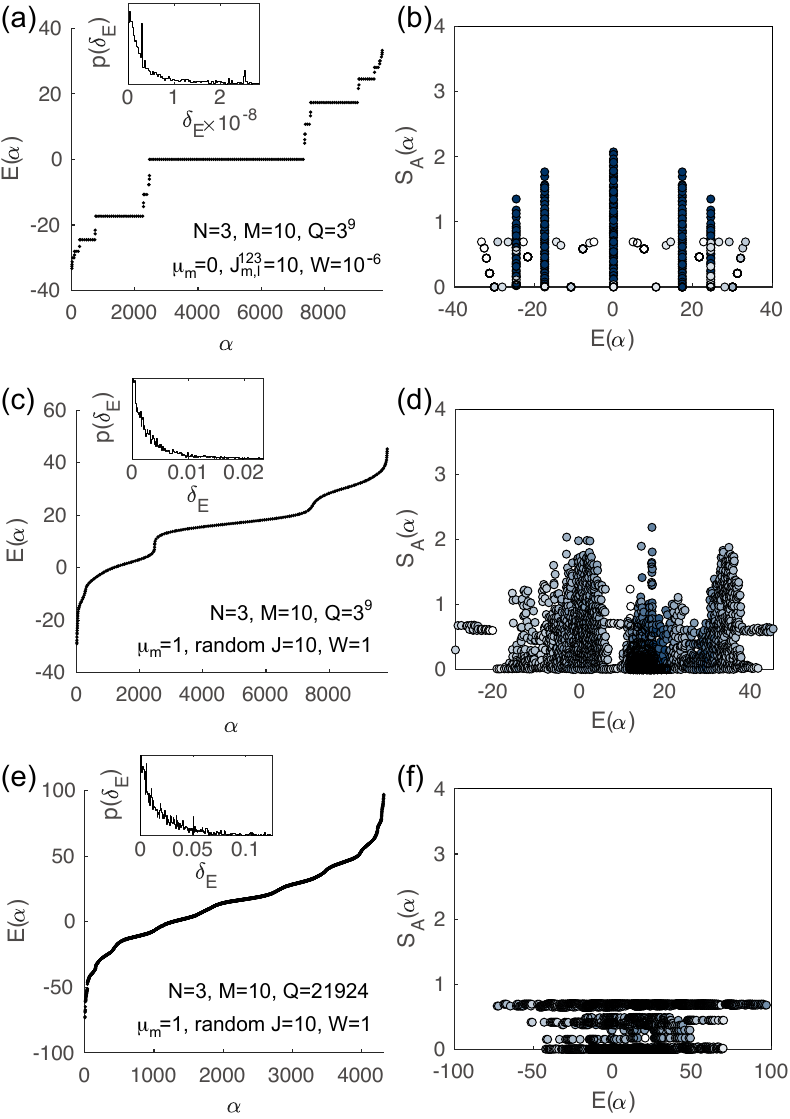}
\end{center}
\caption{ED calculation in a charge $Q$ sector of the model with $(N,M)=(3,10)$ and disorder potential $W>0$. The parameters except for $W$ are exactly the same as \cref{figN=3-W=0} and labeled in the left panels. (a)-(b) has $W=10^{-6}$, while (c)-(f) has $W=1$. (a), (c), (e) shows the energy levels $E(\alpha)$, and their LSS in the insets. (b), (d), (f) show the eigenstate entanglement entropy in subregion $A$ (the first $\lfloor M/2\rfloor$ sites) for (a), (c), (e), respectively (the darker the higher density of dots).
}
\label{figN=3-W>0}
\end{figure}

The MBL nature of the eigenstates of the $N=3$ quantum breakdown model can be strengthened when the disorder potential $H_{\text{dis}}$ in \cref{eq:model-Hd} is turned on, namely, when the disorder strength $W>0$ (defined in \cref{eq-W}).

In the presence of a nonzero $H_{\text{dis}}$ term, the basis transformation in \cref{eq:trans-H3} can no longer simplify the model (there is no Krylov subspace structure), and there is no obvious analytical method for solving the eigenstates. We therefore calculate the eigenstates and spectrum of a charge $Q$ sector numerically by ED.

Fig. \ref{figN=3-W>0} shows three examples of the ED energy spectrum and subregion entanglement entropy of the model with $(N,M)=(3,10)$, where the model parameters except for $W$ are exactly the same as those in the corresponding panels of Fig. \ref{figN=3-W=0}. Generically, the disorder potential $W$ further localizes the eigenstates and strengthen the MBL, as can be seen from the lower entanglement entropy of eigenstates in Fig. \ref{figN=3-W>0}(b),(d),(e) as compared to Fig. \ref{figN=3-W=0}(b),(d),(e) (where $W=0$). We discuss this result in more details below.

First, a tiny disorder potential can readily pin the originally degenerate many-body eigenstates into a more local basis with considerably lower entanglement entropy. Fig. \ref{figN=3-W>0}(a)-(b) shows the numerical calculation in the $Q=3^9$ sector by adding a disorder potential strength $W=10^{-6}$ to the $(N,M)=(3,10)$ case with parameters in Fig. \ref{figN=3-W=0}(a)-(b), after which the energy spectrum is almost unchanged (see Figs. \ref{figN=3-W=0}(a) and \ref{figN=3-W>0}(a)). The inset of Fig. \ref{figN=3-W>0}(a) shows the LSS of in the range of $\delta_E\sim 10^{-8}$ (the energy scale of level degeneracy breaking), which is still a Poisson distribution. In contrast, the entanglement entropies of the degenerate levels in Fig. \ref{figN=3-W>0}(b) are much lower than those in Fig. \ref{figN=3-W=0}(b), indicating the eigenstates are more localized. This is because the disorder potential $W$ connects the Krylov subspaces with different blocking sites $m_b$ and different occupation configurations $\{s^j_m\}$, which induces a localization and degeneracy breaking among all the originally degenerate eigenstates in different Krylov subspaces, similar to the Anderson localization.

The energy level degeneracy breaking and MBL become stronger as the disorder strength $W$ increases. Fig. \ref{figN=3-W>0}(c)-(f) show the spectra in the $Q=3^9$ and $Q=21924$ charge sectors at $(N,M)=(3,10)$ with $W=1$, where the other parameters are exactly the same as those in Fig. \ref{figN=3-W=0}(c)-(f). Compared with Fig. \ref{figN=3-W=0}(c)-(f), the originally degenerate levels at distinct energies are broadened and merged together by the disorder potential $W$, showing a clear Poisson LSS in Fig. \ref{figN=3-W>0}. As shown in Fig. \ref{figN=3-W>0}(d) and (f), extensive numbers of eigenstates show extremely low entanglement entropy regardless of energy, indicating MBL. The $Q=21924$ sector (Fig. \ref{figN=3-W>0}(f), as a representative of generic charge sector) is clearly much more localized than the $Q=3^9$ sector (Fig. \ref{figN=3-W>0}(d)), due to the breaking of the system into smaller decoupled or weakly coupled subsystems as explained in Sec. \ref{sec:N=3-generic-Q-approx}. In general, we find the MBL is robust for any value $W>0$. The \cref{app:N>3} \cref{figS-N>3}(a)-(b) shows an example of the $N=3$ model with $W=10$ comparable to the interaction strength $J=10$, where the MBL feature is still clear. In the $W\gg J$ limit, the MBL is even stronger, as the disorder potential $W$ also favors the localization of fermions.

%As mentioned earlier, the disorder potential $W$ can be understood as disorder of the ionization energy of atoms in the dielectric gas breakdown.

\section{The $N>3$ models}\label{sec:N>3}

The quantum breakdown model in \cref{eq:model} with $N>3$ becomes increasingly complicated, and we have not found a method to solve it analytically. Therefore, we employ ED to numerically compute the full energy spectrum and eigenstates of the $N>3$ model. Within the system sizes (up to $NM\approx 25\sim 28$) and charge $Q$ sectors calculable, when the disorder potential $W$ is not strong ($W<J$), we find the model exhibits a crossover from MBL to quantum chaotic behaviors as $M/N$ decreases to $M/N\approx 1$ (see subsection \ref{sec:NMtrans} below). This suggests that the model has an MBL localization length of order $N$ sites, and behave like a chaotic quantum dot within a localization length. Therefore, we conjecture the model is in a 1D MBL phase with localization length $M_L\sim N$ in the $M/N\rightarrow \infty$ limit (the 1D limit), and is many-body chaotic in the $M/N\rightarrow 0$ limit (the 3D limit). At strong disorder potentials $W\gg J$, we find the model always shows MBL features irrespective of $M/N$ (see subsection \ref{sec:MBLstrongW}).

Moreover, for generic $(N,M)$ at disorder potential $W=0$, we can identify a set of exactly solvable degenerate many-body eigenstates in many charge $Q$ sectors, which form a flat band of many-body quantum scar states. This is discussed in subsection \ref{sec:scar-band}. 

%Within the limited system sizes and charge $Q$ sectors numerically accessible, we find the model shows crossover features between MBL and quantum chaos with quantum scar states when $M/N>1$, and is strongly quantum chaotic with quantum scar states when $M/N<1$. This suggests that the model effectively form localized quantum chaotic dots with a localization length of order $N$ number of sites, and thus we conjecture the model is always in a 1D MBL phase with localization length of order $N$ in the $M/N\rightarrow \infty$ limit. The details are discussed in subsection \ref{sec:NMtrans} below.

\subsection{MBL to chaos crossover as $M/N$ decreases at small $W$}\label{sec:NMtrans}

Similar to the $N=3$ case, we first discuss the $N>3$ quantum breakdown model at disorder potential $W=0$, and then examine the robustness of the model features at small disorder potential $W>0$ ($W< J$). In the ED calculations, we set a uniform chemical potential $\mu_m=\mu$, and random interactions with a strength $J$, as we assumed in \cref{eq-muJ,eq-J}. Throughout this section, we set $\mu=1$ and $J=10$.

\subsubsection{ED result at zero and small $W$}

\begin{figure}[tbp]
\begin{center}
\includegraphics[width=3.3in]{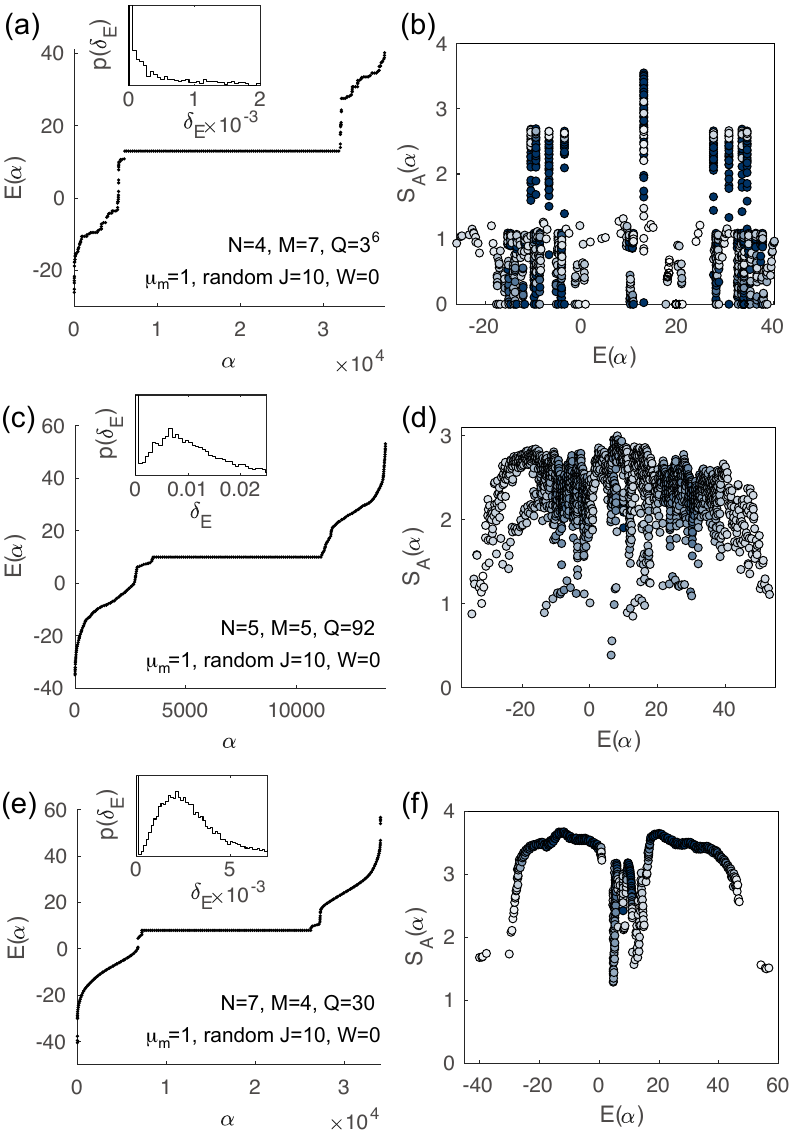}
\end{center}
\caption{ED calculation in a charge $Q$ sector of the models with $N=4,5$ and $7$ and the largest $M$ calculable, respectively, and disorder potential $W=0$. The parameters are labeled in the left panels. (a), (c), (e) shows the energy levels $E(\alpha)$, and insets show their LSS. (b), (d), (f) show the eigenstate entanglement entropy in subregion $A$ (the first $\lfloor M/2\rfloor$ sites) for (a), (c), (e), respectively (the darker the higher density of dots).
}
\label{figN>3-W=0}
\end{figure}

Fig. \ref{figN>3-W=0} shows the many-body energy spectrum and entanglement entropy of a representative charge $Q$ sector in the $(N,M)=(4,7)$, $(5,5)$ and $(7,4)$ quantum breakdown model with $W=0$, respectively. The most obvious feature we observe is the presence of an interval of flat dispersion in the middle of the energy spectrum in Fig. \ref{figN>3-W=0}(a),(c),(e), which indicates a highly degenerate energy level in each of the charge $Q$ sectors. Such a degenerate level when $W=0$ is observed in many charge $Q$ sectors at generic $(N,M)$ in our ED. We call this degenerate level a \emph{many-body scar flat band}, the origin of which is explained later in subsection \ref{sec:scar-band}. For $N=3$, this is the level with the highest degeneracy in Fig. \ref{figN=3-W=0}. In addition, in the $(N,M)=(4,7)$ spectrum in Fig. \ref{figN>3-W=0}(a), other than the many-body scar flat band (the highest degeneracy level in the middle), there are enormous other degenerate levels showing a fractal-like structure (similar to \cref{figN=3-W=0}), which cannot be understood by the explanation later in subsection \ref{sec:scar-band}. This suggests the existence of Hilbert space fragmentation into degenerate Krylov subspaces at least also at $N=4$, which we leave for future study.

The insets of Fig. \ref{figN>3-W=0}(a),(c),(e) show the LSS of the energy spectrum of the corresponding charge $Q$ sector. First, all the LSS show a delta function peak at $\delta_E=0$, which is due to the many-body scar flat band and other degenerate levels if any. At $\delta_E>0$, the LSS is Poisson in Fig. \ref{figN>3-W=0}(a) when $(N,M)=(4,7)$, almost Wigner-Dyson when $(N,M)=(5,5)$, and clearly Wigner-Dyson in Fig. \ref{figN>3-W=0}(e) when $(N,M)=(7,4)$. The eigenstate entanglement entropy of Fig. \ref{figN>3-W=0}(a),(c),(e) in subregion A (the first $\lfloor M/2\rfloor$ sites) are given in Fig. \ref{figN>3-W=0}(b),(d),(f), respectively. In Fig. \ref{figN>3-W=0}(b) where $(N,M)=(4,7)$, most eigenstates have low (area law) entanglement entropies strongly fluctuating with energies, and the pattern resembles that of $(N,M)=(3,10)$ in Fig. \ref{figN=3-W=0}, showing MBL features. In contrast in Fig. \ref{figN>3-W=0}(f) where $(N,M)=(7,4)$, most eigenstates have high entanglement entropies, outlining a volume law dome expected for quantum chaotic systems. In Fig. \ref{figN>3-W=0}(d) where $(N,M)=(5,5)$, the entanglement entropies show features in between the above two cases. These results suggest there is a crossover from MBL behavior to quantum chaos behavior as the ratio $M/N$ decreases to around $M/N\approx1$.

\begin{figure}[tbp]
\begin{center}
\includegraphics[width=3.3in]{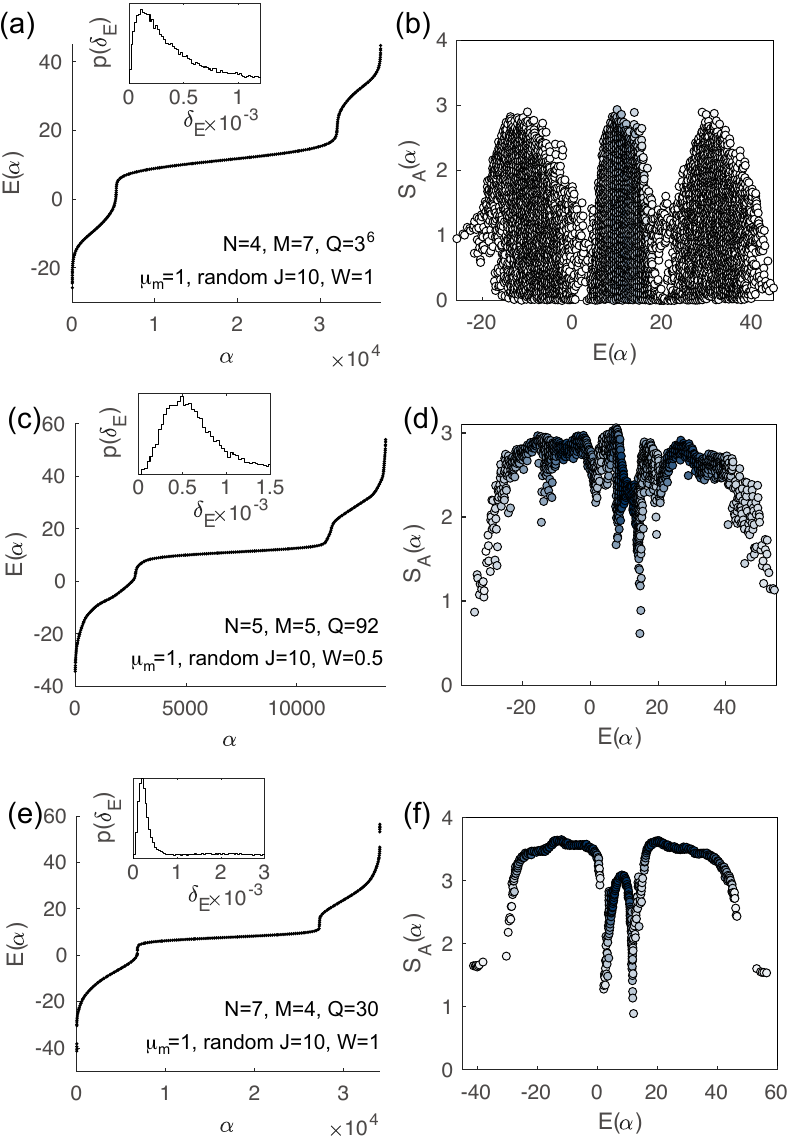}
\end{center}
\caption{ED calculation in a charge $Q$ sector of the models with $N=4,5$ and $7$ and the largest $M$ calculable, respectively, and disorder potential $W>0$. The parameters except for $W$ are the same as \cref{figN>3-W=0}, which are labeled in the left panels. (a), (c), (e) shows the energy levels $E(\alpha)$, and insets show their LSS. (b), (d), (f) show the eigenstate entanglement entropy in subregion $A$ (the first $\lfloor M/2\rfloor$ sites) for (a), (c), (e), respectively (the darker the higher density of dots).
}
\label{figN>3-W>0}
\end{figure}

This crossover behavior remains robust against a small disorder potential $W>0$ ($W<J$, see subsection \ref{sec:MBLstrongW} for the criterion of small $W$). Fig. \ref{figN>3-W>0} shows calculations with a small disorder potential $W>0$ (equals $0.5$ or $1$, see parameters in the panels), where all the other parameters are exactly the same as Fig. \ref{figN>3-W=0}. The degenerate scar level is lifted by $W>0$. When $(N,M)=(4,7)$, most eigenstates still show low (area law) entanglement entropies as shown in Fig. \ref{figN>3-W>0}(b), which is similar to the $N=3$ case in Fig. \ref{figN=3-W>0}(d) and suggests MBL. The LSS in Fig. \ref{figN>3-W>0}(a) inset becomes semi-Poisson with a weak level repulsion. We conjecture this LSS will tend to Poisson if we could calculate larger $M/N$ (when $M$ is larger than the MBL localization length), similar to the $(N,M)=(3,10)$ case in Fig. \ref{figN=3-W>0}(c) inset. This MBL-like behavior is robust for any $W>0$. An example of the $(N,M)=(4,7)$ model with $W=10$ (comparable to $J=10$) is shown in \cref{app:N>3} \cref{figS-N>3}(c)-(d), which show clear MBL features of Poisson LSS and area law eigenstate entanglement entropies. The $(N,M)=(5,5)$ and $(7,4)$ cases with a small $W>0$ (Fig. \ref{figN>3-W>0}(c)-(f)) are not too different from those with $W=0$ (Fig. \ref{figN>3-W=0}(c)-(f)), except that the eigenstate entanglement entropies vary more smoothly with respect to the energy, showing clearer features of volume law behaviors. Note that the LSS in the energy scale of degeneracy breaking of the many-body scar flat band (insets of Fig. \ref{figN>3-W>0}(c),(e)) are also Wigner-Dyson, and this is because the disorder potential $W$ hybridizes the flat band with the other delocalized eigenstates, leading to a level repulsion. Examples of more charge $Q$ sectors with $W=0$ or $W>0$ for the same $(N,M)$ can be found in the App. \ref{app:N>3} Fig. \ref{figS-N>3}. The LSS for smaller $(N,M)$ are shown in App. \ref{app:N>3} Fig. \ref{figS-smallNM}. All of these results suggest the existence of an MBL to chaos behavior crossover as $M/N$ decreases to around $M/N\approx1$.

Moreover, in the $(N,M)=(7,4)$ quantum chaotic case in Fig. \ref{figN>3-W=0}(f) and Fig. \ref{figN>3-W>0}(f), there are sharp dips of entanglement entropy in the middle of the energy spectrum, which indicate these states are low-entanglement quantum scar states. The dips are right above and below the energy of the many-body scar flat band in Fig. \ref{figN>3-W=0}(e), where the density of states are low. Such quantum scar states also occur in the $(N,M)=(5,5)$ case in Fig. \ref{figN>3-W=0}(d) and Fig. \ref{figN>3-W>0}(d). At $W=0$, the many-body scar flat band states in Fig. \ref{figN>3-W=0}(e) can also be understood as low-entanglement quantum scar states, since they can be understood as states of fermions which do not hop, as we will show later in \cref{sec:scar-band}. Adding a tiny disorder $W>0$ will fix the eigenbasis of the flat band into a relatively low entanglement entropy basis (see App. \ref{app:N>3} Fig. \ref{figS-N>3-smallW}), although the $W$ induced hybridization between the flat band and the other delocalized eigenstates will eventually delocalize these flat band eigenstates at larger $W$. In Fig. \ref{figN>3-W=0}(f) with $W=0$, the many-body scar flat band states do not show low entanglement entropy, again because the ED calculation arbitrarily picks a high entanglement entropy eigenbasis within the degenerate Hilbert space. Remarkably, the fraction of such flat band quantum scar eigenstates is not measure zero compared to the total number of eigenstates (see subsection \ref{sec:scar-band}). This is in stark contrast to previous quantum scar models, where the number of scar states is usually measure zero in the thermodynamic limit.

\subsubsection{Argument for the crossover from MBL to chaos}

To define the MBL and quantum chaos phases, certain thermodynamic limit need be taken. As we explained in \cref{sec:model-physical}, we define $M/N\rightarrow\infty$ as the 1D thermodynamic limit (where $N$ can be a small number), and $M/N\rightarrow 0$ as the 3D thermodynamic limit (where both $M$ and $N$ are large). Our small size calculations then suggest the model is MBL in the 1D thermodynamic limit, and quantum chaotic in the 3D thermodynamic limit, and the crossover possibly happens at $M/N\approx1$. We emphasize that this conclusion may be subject to finite size effects. Nonetheless, we provide here a theoretical argument for the crossover around $M/N\approx1$ at small disorder potential $W$.

The key is to identify the localization length $M_{L}$ (number of sites) of the model. Recall that a fermion on the $m$-th site can move to the $(m+1)$-th site and turn into three fermions. If two of these three fermions stay in the $(m+1)$-th site, and the other fermion further move and turn into three fermions on the $(m+2)$-th site (as is the example in \cref{eq-config-N=3}), it appears this process can go on to any distance away without localization. However, we argue this is not true. This is because in this picture, only one fermion in the last occupied site is mobile (subject to randomness induced localization), while the mobilities of all the other fermions rely on this single fermion far way, which would not be a bulk property. To define the localization length $M_L$ as a bulk property, we need to allow at least one fermion per site to be mobile independently within a localization length $M_L$. Since a fermion needs to produce at least two more fermions per site to move forward, it requires a Hilbert space of $2\cdot(M_L/2)=M_L$ fermion modes if it is to move by $M_L/2$ sites, the mid-value of a localization interval $[0,M_L]$. Within an interval of $M_L$ sites, to allow one mobile fermion per site, we would need a Hilbert space of $M_L\cdot M_L=M_L^2$ number of fermion modes; while the total number of fermion modes available in the interval is $NM_L$. Equalizing these two numbers yields a MBL localization length
\begin{equation}\label{eq-loc-ML}
M_L\sim N\ .
\end{equation}
Therefore, we expect the model to show MBL features when $M>M_L$, or $M/N>1$. In contrast, when $M<M_L$, or $M/N<1$, the model should behave as a random extended system, which is usually quantum chaotic.

\subsection{Persistent MBL at $W\gg J$}\label{sec:MBLstrongW}

\begin{figure}[tbp]
\begin{center}
\includegraphics[width=3.3in]{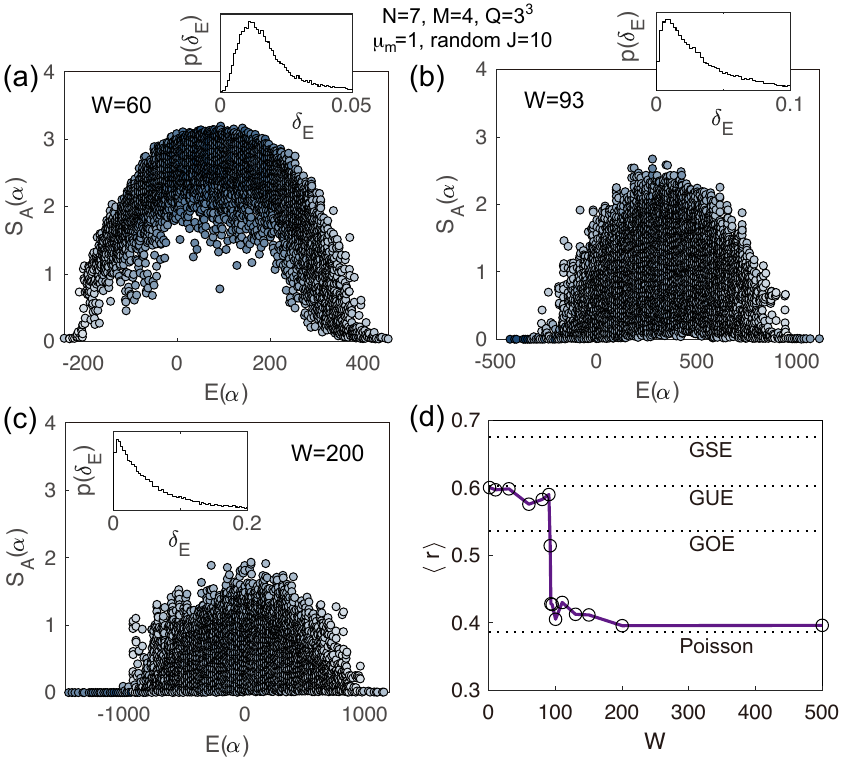}
\end{center}
\caption{The crossover from quantum chaos to MBL with respect to disorder potential strength $W$ in the $(N,M)=(7,4)$ model. The interaction is random with fixed strength $J=10$, and $\mu_m=1$, and the charge $Q=3^3$. (a)-(c) gives the entanglement entropy $S_A(\alpha)$ of eigenstates with energy $E(\alpha)$ in the first $2$ sites, and their LSS (inserts), for disorder potential strength $W=60, 93$ and $200$, respectively. (d) shows the level spacing ratio $\langle r\rangle$ of the model with respect to the disorder potential strength $W$, which shows a crossover from Wigner-Dyson (GUE) to Poisson at $W\approx93$. The 4 horizontal dashed lines give the expected $\langle r\rangle$ values for Poisson ($\langle r\rangle=0.39$), GOE ($\langle r\rangle=0.53$), GUE  ($\langle r\rangle=0.60$) and GSE  ($\langle r\rangle=0.67$), respectively.
}
\label{fig-Ne7-trans}
\end{figure}

We now show that when the disorder potential $W\gg J$, the model exhibits MBL features irrespective of $M/N$. 

When $M/N>1$, we have shown in subsection \ref{sec:NMtrans} that the model shows features of MBL at small disorder potential $W<J$. In this case, increasing the disorder potential $W$ only strengthens the MBL features, as shown in the appendix \cref{figS-N>3}(a)-(d) where $W=J=10$. This means strong disorder potential $W$ will further decrease the localization length $M_L$ in \cref{eq-loc-ML}. Our ED calculation strongly suggests that the model behaves as MBL for any disorder potential strength $W$ when $M/N>1$.

When $M/N<1$, our ED calculation indicates that there is a crossover (or transition) from quantum chaos to MBL as $W/J$ increases. This should correspond to the point where the localization length $M_L$ decreases to $M$. In the case of $(N,M)=(7,4)$ with $\mu_m=1$ and $J=10$, \cref{fig-Ne7-trans}(a)-(c) shows the eigenstate entanglement entropies $S_A(\alpha)$ and the LSS (the insets) in the charge $Q=3^{M-1}$ sector, at different values of $W$. As $W$ increases, the entanglement entropies $S_A(\alpha)$ decrease monotonically, suggesting the approach to the area law, and the LSS crossovers from Wigner-Dyson (GUE) in \cref{fig-Ne7-trans}(a) to Poisson in \cref{fig-Ne7-trans}(c). These features are clear indications of MBL. To reveal this crossover more explicitly, we calculate the mean \emph{level spacing ratio} $\langle r\rangle$ defined as the mean value of the following ratio \cite{oganesyan2007}:
\begin{equation}
r(\alpha)=\frac{\min\{\delta_E(\alpha),\delta_E(\alpha+1)\}}{\max\{\delta_E(\alpha),\delta_E(\alpha+1)\}}\ ,
\end{equation}
where $\delta_E(\alpha)$ is the nearest neighboring level spacing in the charge $Q$ sector defined in \cref{eq:spacing}. For LSS showing Poisson, GOE, GUE and GSE, the level spacing ratio $\langle r\rangle$ is expected to be $0.39$, $0.53$, $0.60$ and $0.67$, respectively \cite{atas2013}. As shown in \cref{fig-Ne7-trans}(d), we find the level spacing ratio in this charge $Q=3^{M-1}$ sector shows a sharp crossover (transition) from the GUE value to the Poisson value at $W/J\approx 9.3$, which is an indication of crossover (or transition) from quantum chaos to MBL as $W/J$ increases.

Therefore, we conclude that at sufficiently strong disorder potential $W\gg J$, the model is always in the MBL regime. Intuitively, the on-site disorder potential may play a similar role as the disorder potential in the Anderson localization problem.

\subsection{Many-body scar flat band at $W=0$}\label{sec:scar-band}

We have seen in Fig. \ref{figN>3-W=0} (see also App. \ref{app:N>3} Fig. \ref{figS-N>3}) that at $W=0$, many charge $Q$ sectors of the quantum breakdown model have an extensively degenerate level in the middle of the spectrum, which we name as the \emph{many-body scar flat band}. Moreover, we find all the eigenstates $|\alpha\rangle$ in this many-body scar flat band has integer number of fermions on each site:
\begin{equation}\label{eq:scar-nm}
\langle \alpha|\hat{n}_m|\alpha\rangle=n_m\in\mathbb{Z}\ ,\quad (n_m\ge0)
\end{equation}
and their eigen-energy is given by
\begin{equation}\label{eq:scar-energy}
E(\alpha)=\sum_{m=1}^M \mu_m n_m\ .
\end{equation}
This indicates that the interaction $H_I$ in Eq. \ref{eq-HI} has no contribution to the eigen-energy of these states at all, so they are the zero modes of $H_I$ satisfying $H_I|\alpha\rangle=0$. We note that \cref{eq:scar-nm,eq:scar-energy} are generically not true for energy levels not in the many-body scar flat band (for instance, not true for any degenerate levels other than the many-body scar flat band at energy $E=13$ in the $(N,M)=(4,7)$ spectrum in Fig. \ref{figN>3-W=0}(a)).
%The energy spectra of a few example charge sectors in $N>3$ models with $W=0$ are as shown in Fig. \ref{figN>3-W=0} and Fig. \ref{figS-N>3}. Except for Fig. \ref{figS-N>3}(a) which is a rare charge sector without such a many-body scar flat band (particularly chosen and shown here), the other figures all have the many-body scar flat band which can be clearly seen in the middle of the energy spectra.

\begin{figure}[tbp]
\begin{center}
\includegraphics[width=3.3in]{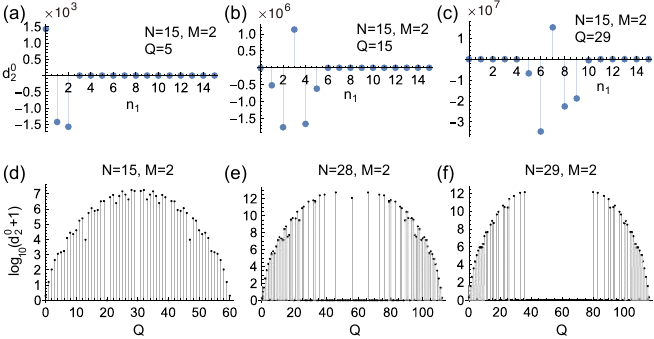}
\end{center}
\caption{(a)-(c) The classical degeneracy $d_2^0(n_1,n_2)$ (defined in \cref{eq:d20}) for several charge $Q$ sectors $(N,M)=(15,2)$ as a function of $n_1$ (recall that $n_2=Q-3n_1$). Each $Q$ sector has only one $(n_1,n_2)$ configuration with $d_2^0(n_1,n_2)>0$. (d)-(f) The maximal $d_2^0$ of each charge $Q$ sector, calculated for $(N,M)=(15,2), (28,2)$ and $(29,2)$, respectively. The $y$ axis is $\log_{10}(d_2^0+1)$ (such that zero corresponds to $d_2^0=0$).
}
\label{figM=2}
\end{figure}

The origin of such a many-body scar flat band here resembles that of the single-particle flat bands of line-graph tight-binding models \cite{mielke1991,dumitru2021} (e.g., the Lieb lattice \cite{lieb1989} and the kagome lattice \cite{bergman2008}), which is due to the mismatch of dimensions of the Hilbert subspaces connected by the Hamiltonian $H_I$, as we will show below.

\subsubsection{$M=2$ sites}\label{sec:scar-band-M=2}

The simplest example is the quantum breakdown model with $M=2$ sites. Consider the Hilbert space of Fock states with $n_m$ ($n_m\in\mathbb{Z}$) fermions on site $m$ ($m=1, 2$), which has a Hilbert space dimension
\begin{equation}\label{eq:hn1n2}
h_{n_1,n_2}=C_{N}^{n_1}C_{N}^{n_2}\ ,
\end{equation}
where $C_N^n=\frac{N!}{(N-n)!n!}$ is the binomial coefficient. Note that $C_N^n=0$ if $n<0$ or $n>N$. The conserved charge $Q$ is given by $Q=3n_1+n_2$. The interaction $H_I$ in \cref{eq-HI} can hop from the Hilbert space of fermion numbers $\{n_1,n_2\}$ into the Hilbert space of fermion numbers $\{n_1\pm1,n_2\mp 3\}$. The $H_I$ hopping matrix between these two Hilbert spaces thus has $h_{n_1+1,n_2-3}+h_{n_1-1,n_2+3}$ rows and $h_{n_1,n_2}$ columns. If the Hilbert space dimensions for some integers $\{n_1,n_2\}$ satisfy
\begin{equation}\label{eq:d20}
d_2^0(n_1,n_2)=h_{n_1,n_2}-h_{n_1+1,n_2-3}-h_{n_1-1,n_2+3}>0\ ,
\end{equation}
the above $H_I$ hopping matrix will have a rank no larger than $h_{n_1+1,n_2-3}+h_{n_1-1,n_2+3}$, and thus there will necessarily be
\begin{equation}
d_2(n_1,n_2)\ge d_2^0(n_1,n_2)
\end{equation}
number of linearly independent states in the $\{n_1,n_2\}$ Hilbert space that can be annihilated by $H_I$. Namely, the interaction Hamiltonian $H_I$ will have $d_2$ zero modes with fermion numbers $\{n_1,n_2\}$, which will be eigenstates with energy given by \cref{eq:scar-energy} when the disorder potential $W=0$. These zero modes of $H_I$ therefore form a many-body scar flat band. We call $d_2^0(n_1,n_2)$ in Eq. \ref{eq:d20} the \emph{classical degeneracy} (from simple dimension counting), and the actual degeneracy $d_2(n_1,n_2)$ the \emph{quantum degeneracy}.

Fig. \ref{figM=2}(a)-(c) shows the classical degeneracy $d_2^0(n_1,n_2)$ defined in \cref{eq:d20} as a function of $n_1$ in several charge $Q$ sectors for $(N,M)=(15,2)$ (recall $n_2=Q-3n_1$). Each of them has one configuration $\{n_1,n_2\}$ with $d_2^0(n_1,n_2)>0$, and thus has a many-body scar flat band. Generically, for $N\le28$, most charge $Q$ sectors have one (and usually only one) fermion number configuration $\{n_1,n_2\}$ with $d_2^0(n_1,n_2)>0$, as shown in Fig. \ref{figM=2}(d)-(e). When $N\ge 29$, a charge $Q$ sector can have $d_2^0(n_1,n_2)>0$ only if $Q\lesssim N^{2/3}$ or $4N-Q\lesssim N^{2/3}$, and the example of $N=29$ is shown in Fig. \ref{figM=2}(f). We give a theoretical understanding and estimation of such behaviors with respect to $N$ in App. \ref{app:zero-modes}.

Within the ED calculable system sizes with $M=2$, we find $d_2=d_2^0$ in all charge sectors except for one case: the $Q=2N$ sector when $N$ is odd. In this case, within $N\le7$, we find $d_2-d_2^0\le2$. Therefore, the classical degeneracy $d_2^0$ gives a very accurate estimation of the quantum degeneracy $d_2$.

\subsubsection{Generic $M$ sites}

The counting of many-body scar flat band degeneracy in Eq. \ref{eq:d20} for $M=2$ sites can be straightforwardly generalized to $M>2$ sites. Consider a fermion number configuration $\{n_1,n_2,\cdots,n_M\}$ on the $M$ sites in a charge sector $Q$. We assume the first $m$ sites ($1\le m\le M$) with the sub-configuration $\{n_1,n_2,\cdots,n_m\}$ has $d_M^{(m)}$ zero modes of $H_I$. When we add the $m$-th site to the first $m-1$ sites, the $d_M^{(m)}$ zero modes of the first $m$ sites should span a subspace of the direct product of the zero-mode Hilbert space of the first $m-1$ sites (dimension $d_M^{(m-1)}$) and the $n_m$-fermion Hilbert space of the $m$-th site (dimension $C_N^{n_m}$). This is because the terms in $H_I$ within the first $m-1$ sites have to annihilate these $d_M^{(m)}$ zero modes. Then, these $d_M^{(m)}$ zero modes also need be annihilated by the interaction $J_{m-1,l}^{ijk}$ between sites $m-1$ and $m$. From our discussion in Sec. \ref{sec:scar-band-M=2}, the matrix of the $J_{m-1,l}^{ijk}$ term within the two sites $m-1$ and $m$ has a rank no larger than $h_{n_{m-1}+1,n_m-3}+h_{n_{m-1}-1,n_m+3}$, where $h_{n_1,n_2}$ is the two-site Hilbert space dimension of fermion number $\{n_1,n_2\}$ defined in \cref{eq:hn1n2}. Therefore, the number of modes that the $J_{m-1,l}^{ijk}$ term cannot annihilate is at most its rank times the number of zero modes $d_M^{(m-2)}$ of the first $m-2$ sites. This indicates the number of zero modes $d_M^{(m)}$ of the first $m$ sites satisfy the recursion relation
\begin{equation}\label{eq:dMm-recursion}
\begin{split}
&d_M^{(m)}=d_M^{(m-1)}C_{N}^{n_m}\\
&-d_M^{(m-2)}(h_{n_{m-1}+1,n_m-3}+h_{n_{m-1}-1,n_m+3})+r_M^{(m)},
\end{split}
\end{equation}
where 
\begin{equation}\label{eq:rMm}
r_M^{(m)}\ge 0
\end{equation}
is a nonnegative quantum correction beyond the above rank counting, which may arise from destructive quantum interferences. The two initial values of the recursion relation are given by 
\begin{equation}
d_M^{(0)}=1\ ,\qquad  d_M^{(1)}=C_{N}^{n_1}\ ,
\end{equation}
where we allowed $m$ to start from $0$ for convenience. The number of zero modes, namely, quantum degeneracy $d_M(\{n_m\})$ of the many-body scar flat band in the total $M$ sites is given by
\begin{equation}
d_M=d_M^{(M)}
\end{equation}
from the recursion relation in \cref{eq:dMm-recursion}.

\begin{figure}[tbp]
\begin{center}
\includegraphics[width=3.3in]{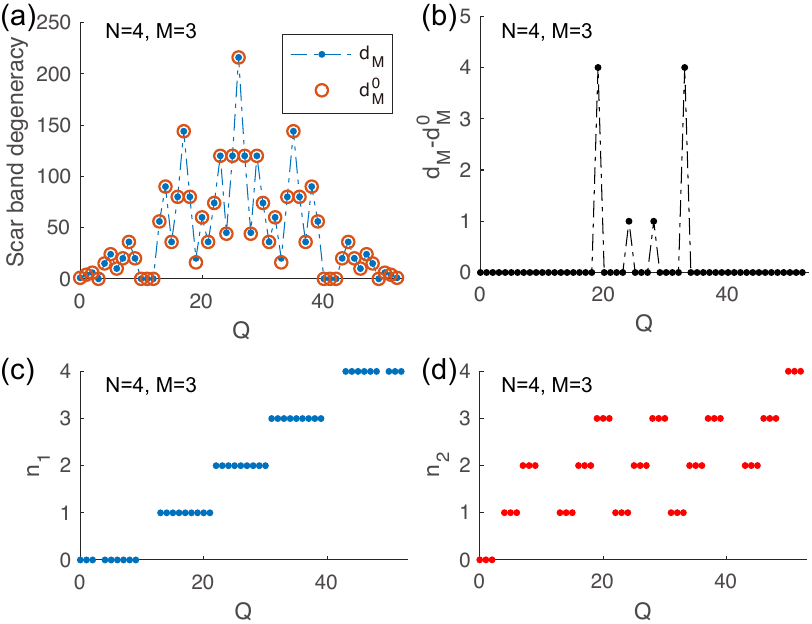}
\end{center}
\caption{Many-body scar flat band for $(N,M)=(4,3)$ with $W=0$ and random interactions. (a) The quantum degeneracy $d_M$ from ED and the classical degeneracy $d_M^0$ computed from \cref{eq:dMm0-recursion} in all charge $Q$ sectors. (b) The difference $d_M-d_M^0$ for all $Q$. (c)-(d) Fermion numbers $n_1$ and $n_2$ in the many-body scar flat band in each charge $Q$ sector. Missing points indicate the absence of a many-body scar flat band.
}
\label{figM=3}
\end{figure}

We have not found a general formula for the quantum correction $r_M^{(m)}$ in \cref{eq:rMm}. From ED calculations, we find $r_M^{(m)}$ is zero in most charge $Q$ sectors. In the rare cases where $r_M^{(m)}>0$, it is always much smaller than $d_M^{(m)}$, so it only gives a small correction to $d_M^{(m)}$. We therefore define a classical degeneracy $d_M^{0}(\{n_m\})$ as
\begin{equation}
d_M^0=d_M^{(M),0}\ ,
\end{equation}
where $d_M^{(M),0}$ is defined by the recursion relation ignoring the quantum correction $r_M^{(m)}$, namely,
\begin{equation}\label{eq:dMm0-recursion}
\begin{split}
&d_M^{(m),0}=d_M^{(m-1),0}C_{N}^{n_m}\\
&\quad -d_M^{(m-2),0}(h_{n_{m-1}+1,n_m-3}+h_{n_{m-1}-1,n_m+3})\ ,
\end{split}
\end{equation}
with initial values $d_M^{(0),0}=1$ and $d_M^{(1),0}=C_{N}^{n_1}$. We expect the classical degeneracy $d_M^0$ not to differ much from the actual quantum degeneracy $d_M$.

Similar to the $M=2$ case in Sec. \ref{sec:scar-band-M=2}, for small $N$ (at least for $N$ calculable in our ED) and $M>2$, most charge sectors $Q$ have one (or very rarely two, see an example in App. \ref{app:M>2} Fig. \ref{figS-DegeM=34}(e)) fermion number configuration $\{n_m\}$ with $d_M>0$, leading to a many-body scar flat band. As an example, Fig. \ref{figM=3}(a) shows the many-body scar flat band quantum degeneracy $d_M$ from ED and classical degeneracy $d_M^0$ from \cref{eq:dMm0-recursion} in all charge sectors $Q$ for $N=4,M=3$, and Fig. \ref{figM=3}(b) shows their difference $d_M-d_M^0$. We see that in most charge $Q$ sectors, $d_M=d_M^0$ precisely. There are only four charge sectors with $d_M-d_M^0>0$, but this difference is much smaller than $d_M$. Fig. \ref{figM=3}(c)-(d) shows the fermion numbers on the first site and second site of the many-body scar flat band, respectively (missing points meaning no many-body scar flat band in that charge $Q$ sector).

%is because all the operators between two neighboring sites in the interaction term $H_I$ are hopping between two many-body Hilbert subspaces with unequal dimensions. Therefore, the rank of the hopping matrix of $H_I$ is upper bounded by the dimension of the smaller Hilbert subspace, leading to zero modes

%\onecolumngrid
%\begin{widetext}
%\begin{figure}[tbp]
%\begin{center}
%\includegraphics[width=6.8in]{figN=4-M=7}
%\end{center}
%\caption{The case of $N=4, M=7$, $Q=3^6$.
%}
%\label{figN=4-M=7}
%\end{figure}

%\begin{figure}[tbp]
%\begin{center}
%\includegraphics[width=6.8in]{figN=5-M=5}
%\end{center}
%\caption{The case of $N=5, M=5$, $Q=3^4$.
%}
%\label{figN=5-M=5}
%\end{figure}

%\twocolumngrid
%\end{widetext}

\section{The breakdown transition}\label{sec-TE}

To examine if the quantum breakdown model can describe the dielectric breakdown phenomena, we numerically calculate the time-evolution of a fermion added into the particle vacuum state of the system. %Under random interactions, the system with chemical potentials $\mu$ and $-\mu$ are related by a particle-hole transformation, so we only study nonnegative chemical potential $\mu\ge0$ here.
We consider uniform chemical potential $\mu_m=\mu$ and random interactions with strength $J=10$. By the chiral symmetry (\cref{sec:chiral}), we can assume $\mu\ge0$ without loss of generality. We will examine both weak and strong disorder potentials $W$.

As explained in \cref{sec:model-physical}, in the dielectric breakdown picture, $\mu$ can be understood as the ionization energy for producing a fermion, with disorder potential within a range of $W$, while $J$ can be viewed as the electric field induced potential energy difference between neighboring layers (the energy a fermion can gain upon moving forward by one site). 
%Physically, the chemical potential $\mu$ can be understood as the ionization energy for producing a fermion. The interaction strength $J>0$ defined in \cref{eq-J} gives the mean root square value of the hopping energy from a one-fermion state in site $m$ to all the possible three-fermion states in site $m+1$, therefore, $J$ can be understood as the average energy an ion gains from the electric field by traveling an interlayer distance $2\ell_{\text{mfp}}$ in Fig. \ref{fig-model}(a). 

In the case of weak disorder potential $W\ll J$, when $J>2\mu$, the energy a fermion gains when moving a site forward will be able to overcome the energy cost of producing two more fermions, allowing the particle avalanche to happen. Therefore, we expect the breakdown transition to happen at
\begin{equation}\label{eq:transition-point}
\frac{\mu}{J}=\frac{1}{2}\ ,\qquad (\mu\ge 0,\quad W\ll J)\ .
\end{equation}
As we will show, this point indeed corresponds to a breakdown transition in the quantum breakdown model with $W\ll J$. %Hereafter, we call $\mu/J>1/2$ the \emph{dielectric regime}, and $\mu/J<1/2$ the \emph{breakdown regime}, respectively.

In the case of strong disorder potential $W\gg J$, when a fermion moves forward by one site, the ionization energy cost of producing two more fermions will be random and roughly within the interval $[2(\mu-W),2(\mu+W)]$. In this case, we expect no breakdown happening, since the system is in the MBL regime (see \cref{sec:MBLstrongW}). However, we expect there is a crossover between weak and strong localization with respect to $\mu/W$. If $J<2(|\mu|-W)$, the interaction energy the fermion gains will not be able to overcome the ionization energy, obstructing the fermion to move by one site forward, so the fermions are strongly localized (almost within one site), which is analogous to strong Wannier-Stark localization. In contrast, if $J>2(|\mu|-W)$, there is a chance the fermion can move to the next site, if the random ionization energy happens to be small, which gives a weaker localization of the fermions analogous to the Anderson localization. Therefore, we expect a crossover from weak to strong localization at
\begin{equation}\label{eq:transition-point-strongW}
\frac{\mu}{W}=1+\frac{J}{2W}\approx 1\ ,\qquad (\mu\ge 0,\quad W\gg J)\ .
\end{equation}
We will show this crossover is also verified by our ED calculations.

\subsection{Energy spectrum dependence on $\mu/J$}

\begin{figure}[tbp]
\begin{center}
\includegraphics[width=3.3in]{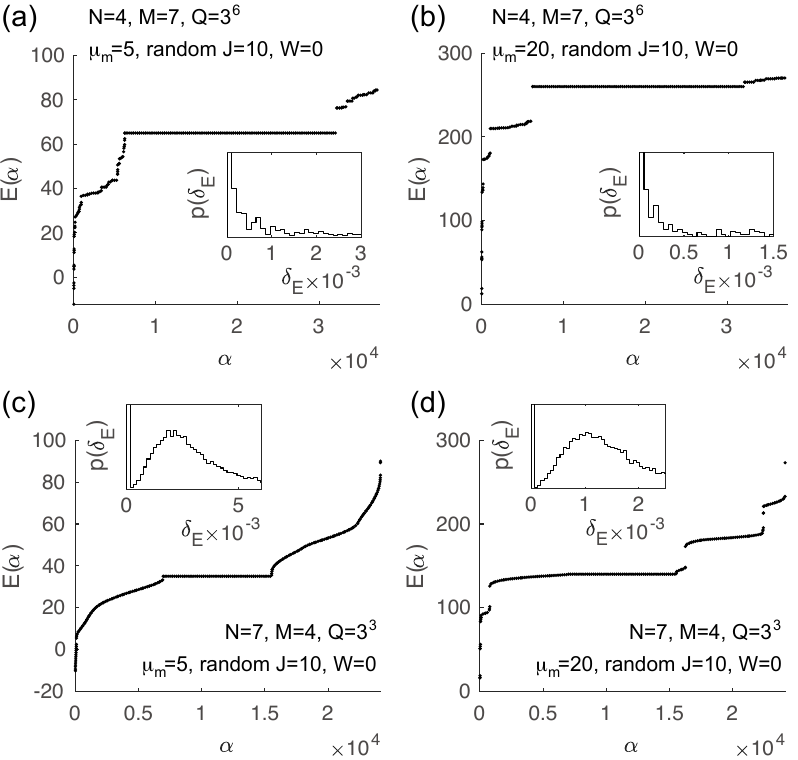}
\end{center}
\caption{The dependence of the energy spectrum on the chemical potential $\mu_m=\mu$, for fixed random interactions with strength $J=10$ and disorder potential $W=0$, and the charge sector is $Q=3^{M-1}$. (a)-(b) has $(N,M)=(4,7)$, while (c)-(d) has $(N,M)=(7,4)$. At large $\mu$, the spectrum splits into bands with gaps around $2\mu$. The LSS has no qualitative dependence on $\mu$.
}
\label{fig-En-mu}
\end{figure}

We first investigate the dependence of many-body energy spectrum on the ratio $\mu/J$. In Figs. \ref{figN=3-W=0} to \ref{figN>3-W>0}, we have shown the calculations only at $\mu/J\leq 0.1$. Fig. \ref{fig-En-mu} shows the energy spectrum with $W=0$ in the $Q=3^{M-1}$ sector at larger $\mu/J$. We find that at small $\mu/J$, the energy spectrum does not have large gaps (Fig. \ref{fig-En-mu}(a),(c)), while at large $\mu/J$, the energy spectrum splits into many-body bands with gaps of order $2\mu$ (Fig. \ref{fig-En-mu}(b),(d)). This is because as $\mu/J\rightarrow\infty$, or $J\rightarrow 0$, the energy spectrum is simply given by $H_\mu$ in \cref{eq:number-operator}, which forms bands with an interband gap $2\mu$. However, we observed no sharp changes but only an adiabatic crossover in the energy spectrum as $\mu/J$ crosses $1/2$. Therefore, if a sharp breakdown transition exists at $\mu/J=1/2$, the change in eigen-wavefunctions must play an essential role. %This is in contrast to a sharp transition can be observed at $\mu/J=\frac{1}{2}$ in time-evolution calculations.

We also find that the LSS of the energy spectrum is insensitive to $\mu/J$. As shown in the insets of Fig. \ref{fig-En-mu}, the LSS of the $(N,M)=(4,7)$ model at $\mu/J=\frac{1}{2}$ and $2$ are both Poisson (Fig. \ref{fig-En-mu}(a)-(b)), while the LSS of the $(N,M)=(7,4)$ model at $\mu/J=\frac{1}{2}$ and $2$ are both Wigner-Dyson (Fig. \ref{fig-En-mu}(c)-(d)). This suggests that the MBL to chaos crossover at $M/N\approx1$ is independent of $\mu/J$.

\subsection{Time evolution}

To reveal the breakdown transition, we examine the dynamical evolution of an initial state $|\psi(0)\rangle$ which adds one fermion to the first site of the particle vacuum $|0\rangle$. Since the interactions are random, we can assume the fermion is in the first fermion mode of site $m=1$ without loss of generality, namely:
\begin{equation}\label{eq:initial-state}
|\psi(0)\rangle=c_{1,1}^\dag |0\rangle\ ,
\end{equation}
where $c_{m,i}^\dag$ is the fermion creation operator in \cref{eq-HI}. Such an initial state resembles creating one ionized atom in an undisturbed dielectric gas. We then employ ED to calculate the time-evolved state $|\psi(t)\rangle=e^{-iHt}|\psi(0)\rangle$ at time $t$. Note that $|\psi(t)\rangle$ is always in the charge $Q=3^{M-1}$ sector. Here we always take random interactions with interaction strength $J=10$, and uniform chemical potential $\mu_m=\mu$. We will examine different disorder potential strength $W$. Note that the wavefunctions $|\psi(t)\rangle$ at chemical potentials $\mu$ and $-\mu$ are related by the chiral symmetry $C$ (\cref{eq:chiral}) followed by complex conjugation, giving the same dynamics. We thus restrict ourselves to $\mu\ge0$.

\begin{figure}[tbp]
\begin{center}
\includegraphics[width=3.3in]{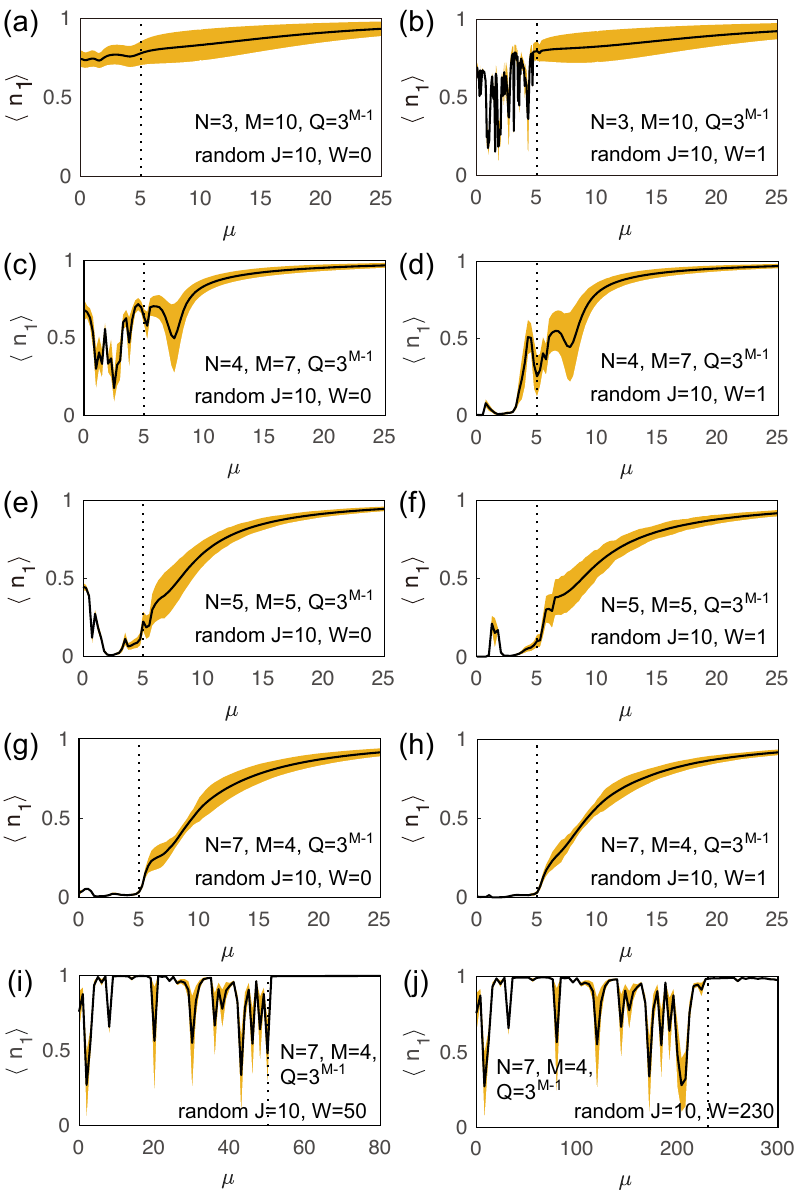}
\end{center}
\caption{The long-time ($t\in[50,100]$) mean value (solid lines) and range of fluctuation (yellow shaded regions) of the number of fermions $n_1$ on site $m=1$ for the initial state in \cref{eq:initial-state}, as a function of chemical potential $\mu_m=\mu$. The parameters are given in the panels. The vertical dashed line in (a)-(h) (for which $W\ll J$) shows the position where $\mu/J=1/2$, while the vertical dashed line in (i)-(j) (for which $W\gg J$) shows the position where $\mu/W=1$.
}
\label{fig-n1}
\end{figure}

We comment that the time evolution of the single fermion state in \cref{eq:initial-state} is highly relevant to the measurement process in a Geiger counter made of dielectric gas in a near-breakdown electric field, where an incident charged particle ionizes an electron from the unperturbed particle vacuum state, giving an initial state similar to \cref{eq:initial-state}. If the breakdown happens, a macroscopic electric current will be arise due to the particle avalanche (\cref{fig-model}(a)). We noticed that most recently, similar measurement-related avalanche phenomenon is studied independently in a spin model with asymmetric interactions in a Bethe-tree lattice \cite{hu2022_measurement,liu2022_measurement}.

We define the number of fermions on the $m$-th site at time $t$ as
\begin{equation}
n_m (t) =\langle \psi(t)|\hat{n}_m|\psi(t)\rangle\ ,
\end{equation}
with the number operator $\hat{n}_m$ defined in \cref{eq:number-operator}. The breakdown can be observed by measuring the number of fermions $n_1(t)$ on the first site. If the breakdown happens, we expect $n_1(t)$ to approach zero as time $t\rightarrow\infty$, since the fermion will turn into many fermions on the $m>1$ sites. In contrast, if the breakdown does not happen, one expects a nonzero $n_1(t)$ as $t\rightarrow\infty$, since the fermion cannot induce a fermion avalanche and is trapped near the first site.

We calculate the time evolution within $t\in[0,100]$, which is sufficient for examining the long time behavior of the evolution. We discuss the weak and strong disorder potential cases separately as follows.

(i) At weak disorder potential $W\ll J$, Fig. \ref{fig-n1} (a)-(h) show the late time behavior of the model with different $(N,M)$ (with $W=0$ in the left panels and $W=1$ in the right panels), where we vary $\mu\in[0,25]$ while fixing $J=10$. The black solid line in each panel shows the long-time mean value $\langle n_1\rangle$ of the fermion number $n_1(t)$ in the late time window $t\in [50,100]$. The yellow shaded region shows the range of fluctuation of $n_1(t)$ in the time window $t\in [50,100]$, which has boundaries at $\langle n_1\rangle\pm\sqrt{\langle \delta n_1^2\rangle}$, where $\sqrt{\langle \delta n_1^2\rangle}$ is the root mean square of the fluctuation $\delta n_1(t)=n_1(t)-\langle n_1\rangle$. The model parameters are given in each panel.

\begin{figure}[tbp]
\begin{center}
\includegraphics[width=3.3in]{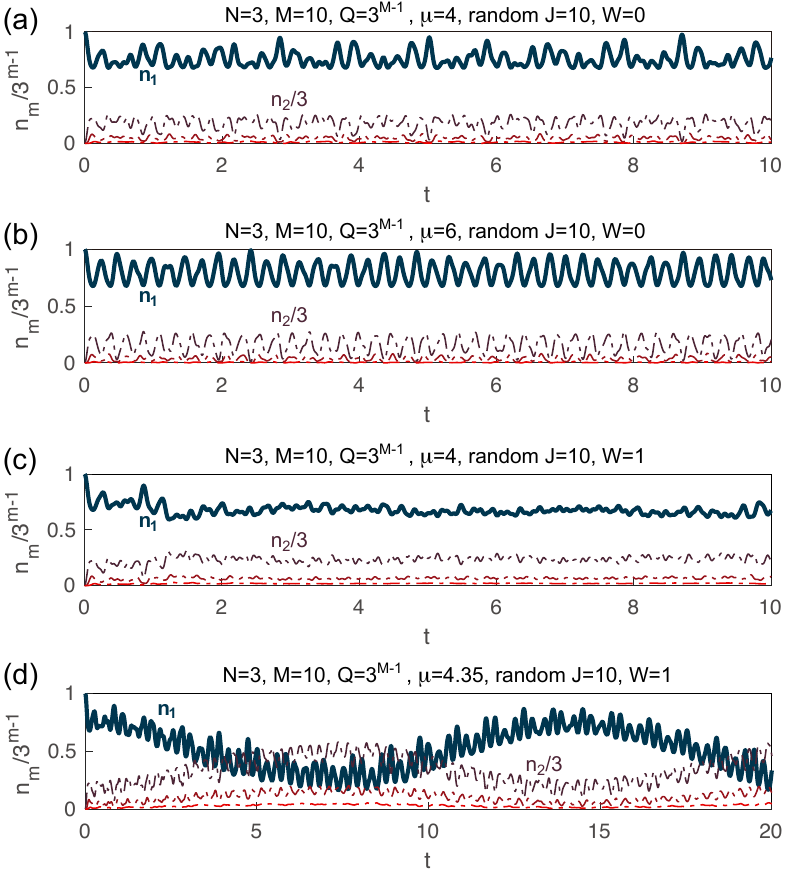}
\end{center}
\caption{The number of fermions $n_m(t)/3^{m-1}$ on site $m$ as a function of time $t$, for $(N,M)=(3,10)$ with the initial state in \cref{eq:initial-state}. $n_1(t)$ is highlighted as the thick solid lines. The parameters are given at the top of each panel (the same $H_I$ and $H_{\text{dis}}$ as \cref{fig-n1}(a),(b)).
}
\label{fig-TE3}
\end{figure}

The most prominent feature is, regardless of the values of $N$ and $M$, the long-time mean value $\langle n_1\rangle$ shows a transition around $\mu=5$, or $\mu/J=1/2$, which is exactly the transition point we expected in \cref{eq:transition-point}. On the $\mu/J>1/2$ side, the mean value $\langle n_1\rangle$ curve is (mostly) smoothly approaching $1$ as $\mu$ grows, with the fluctuation $\sqrt{\langle \delta n_1^2\rangle}$ becoming smaller and smaller. This is the expected dielectric behavior before breakdown, since the fermion is frozen on the first site when $\mu\rightarrow\infty$. The behavior of $\mu/J<1/2$ side depends on $N/M$ (for total fermion modes $NM$ around $25\sim 30$). In Fig. \ref{fig-n1} (a)-(d) where $N/M<1$,  the value $\langle n_1\rangle$ shows oscillations with respect to $\mu$. A special case is $(N,M)=(3,10)$ with disorder potential $W=1$ in Fig. \ref{fig-n1}(b), where $\langle n_1\rangle$ shows a sharp transition from intense oscillation to no oscillation as $\mu/J$ crosses $1/2$. In contrast, in Fig. \ref{fig-n1}(e)-(h) with $N/M\ge1$, $\langle n_1\rangle$ is nearly zero, consistent with our expectation for the breakdown phase. This is the most obvious in the $(N,M)=(7,4)$ case with disorder potential $W=1$ in Fig. \ref{fig-n1}(h), where $\langle n_1\rangle$ show a clear transition from almost zero to nonzero at $\mu/J=1/2$. Hereafter, we call $\mu/J>1/2$ the \emph{dielectric regime}, and $\mu/J<1/2$ the \emph{breakdown regime}.

(ii) At strong disorder potential $W\gg J$, due to MBL irrespective to $N/M$ (\cref{sec:MBLstrongW}), the breakdown transition is absent, and long-time mean value $\langle n_1\rangle$ does not reach $0$ at any $\mu$. However, a sharp crossover of the behavior of $\langle n_1\rangle$ can be observed around $\mu/W=1$, as shown in \cref{fig-n1}(i)-(j), where $(N,M)=(7,4)$ and $W\gg J$. The irregular oscillation of $\langle n_1\rangle$ on the $\mu<W$ side indicates the fermions are only weakly localized around site $m=1$ and can fluctuate, while $\langle n_1\rangle\approx1$ on the $\mu>W$ side indicates the fermion is strongly localized within site $m=1$. Such a crossover between weak  and strong localization is in consistency with our expectation in \cref{eq:transition-point-strongW}. Note that such a behavior is analogous to the $N=3$ model with weak disorder $0<W\ll J$ in \cref{fig-n1}(b), except that the crossover points are different.

\begin{figure}[tbp]
\begin{center}
\includegraphics[width=3.3in]{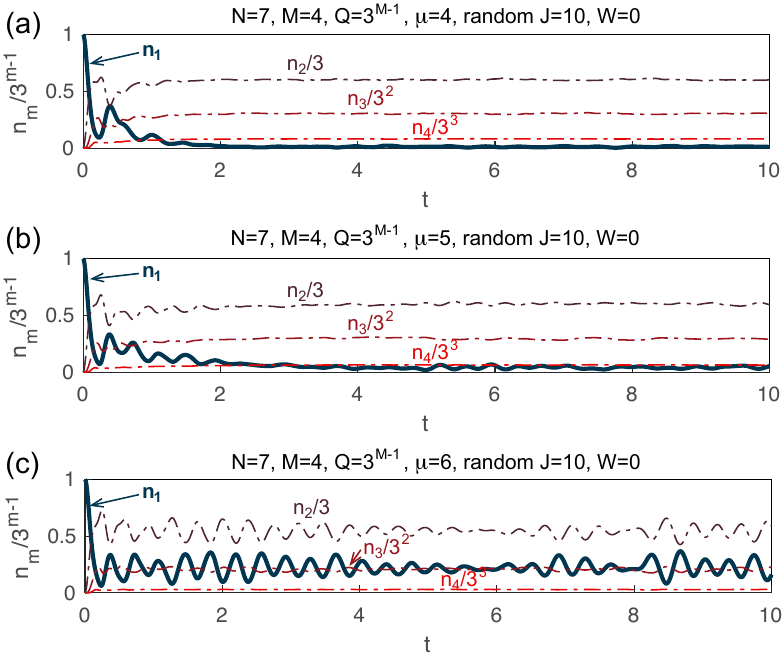}
\end{center}
\caption{The number of fermions $n_m(t)/3^{m-1}$ on site $m$ as a function of time $t$, for $(N,M)=(7,4)$ with the initial state in \cref{eq:initial-state}. $n_1(t)$ is highlighted as the thick solid lines. The parameters are given on top of each panel (the same $H_I$ and $H_{\text{dis}}$ as \cref{fig-n1}(g)).
}
\label{fig-TE}
\end{figure}

We now investigate the time evolutions more closely by examining $n_m(t)$ as a function of time $t$. Since our calculation is in the charge $Q=3^{M-1}$ sector, we have
\begin{equation}
\sum_{m=1}^M \frac{n_m(t)}{3^{m-1}} =\frac{Q}{3^{M-1}}=1\ .
\end{equation}
The $N=3$ case is relatively special compared to the $N>3$ cases, due to its almost analytically solvable nature. Fig. \ref{fig-TE3} shows $n_m(t)/3^{m-1}$ in the $(N,M)=(3,10)$ case, where $n_1(t)$ is highlighted as the thick solid lines. When $W=0$, the difference between $\mu/J<1/2$ and $\mu/J>1/2$ is not significant, both having $\langle n_1\rangle$ far from zero, as Fig. \ref{fig-n1}(a) implies. In both examples of $\mu/J=0.4$ and $\mu/J=0.6$ in Fig. \ref{fig-TE3}(a)-(b), $n_m(t)/3^{m-1}$ show persistent quasi-periodic oscillations with $t$. This can be understood by noting that the initial state in \cref{eq:initial-state} is in either the Krylov subspace without blocking in \cref{eq:sub-Hil-3-0} or the one-dimensional Krylov subspaces in \cref{eq:sub-Hil-3-1} in the $Q=3^{M-1}$ sector of the $N=3$ model with $W=0$, which is equivalent to a tight-binding model with a single effective particle (\cref{eq:H-tb-0}) and thus show no sharp changes as $\mu/J$ varies. When a finite disorder potential $W=1$ is added, the $\mu/J>1/2$ side remains qualitatively unchanged. The $\mu/J<1/2$ side, however, is significantly altered, which develops a strong irregular dependence on chemical potential $\mu$ (see Fig. \ref{fig-n1}(b)). Fig. \ref{fig-TE3}(c) and (d) show two examples with $W=1$ at $\mu/J=0.4$ and $\mu/J=0.435$, respectively. In Fig. \ref{fig-TE3}(c), the oscillation in $n_1(t)$ is suppressed by the disorder potential $W$ (compared to Fig. \ref{fig-TE3}(a)). In Fig. \ref{fig-TE3}(d) (note the range of $t$ plotted is different from Fig. \ref{fig-TE3}(a)-(c)), not only the short period oscillation in $n_1(t)$ is not suppressed, but also a persistent long period oscillation arises. The underlying mechanism of such behaviors remains to be understood, but these oscillations clearly indicate the absence of thermal equilibrium in the $N=3$ model.

The $N>3$ cases do not show significant differences between disorder strength $W=0$ and $0<W\lesssim J$, except that $\langle n_1\rangle$ in the breakdown regime $\mu/J<1/2$ is generically smaller when $0<W\lesssim J$, as Fig. \ref{fig-n1} shows. We therefore only discuss the $W=0$ case here. In Fig. \ref{fig-TE}, we plot the time evolution of $n_m(t)/3^{m-1}$ for $(N,M)=(7,4)$ with $W=0$, where $\mu/J$ is $0.4$, $0.5$ and $0.6$ in Fig. \ref{fig-TE}(a)-(c), respectively. It is clear that $n_1(t)$ decays to zero in the breakdown regime $\mu/J<1/2$, and tends to certain oscillations around a nonzero value in the dielectric regime $\mu/J>1/2$. We note that even if the $(N,M)=(7,4)$ model is quantum chaotic as indicated by its Wigner-Dyson LSS (see Fig. \ref{fig-En-mu}), the dielectric regime $\mu/J>1/2$ shows persistent oscillations in fermion densities and does not thermalize. In the $W\gg J$ case, $\langle n_1\rangle$ shows oscillations depending irregularly on $\mu$ when $\mu<W$ (as \cref{fig-n1}(i)-(j) indicates), similar to the $N=3$ case in \cref{fig-TE3}(c)-(d).

Based on the above observations, we conjecture that, at weak disorder potential $W\ll J$, when both $N$ and $M$ are large enough, irrespective of the ratio $M/N$, the breakdown transition happens robustly at $\mu/J=1/2$, which is signaled by whether $n_1(t)$ tends to zero at large $t$. This is because the breakdown picture only requires sufficiently large number of fermion modes in the subsequent sites. However, we expect the breakdown phase to have a difference between $M/N>1$ and $M/N<1$. When $M/N>1$, our earlier analysis in \cref{sec:NMtrans} implies that the system shows 1D MBL with a localization length around $N$, so we expect the fermion to spread only by a length around $N$ sites in the breakdown regime. Since we expect this MBL to behave as a chaotic quantum dot within a localization length $N$, we expect a local thermal equilibrium within the length $N$ is reached. We call such a phase a \emph{local breakdown}. When $M/N<1$, the model is globally quantum chaotic within a charge $Q$ sector, so we expect the spreading of the fermion to approach a global thermal equilibrium in the breakdown regime, and we call this phase a \emph{global breakdown}. At strong disorder potential $W\gg J$, we expect there is no breakdown for any $M/N$.
%. On the other hand, since the MBL can be understood as coupled local chaotic quantum dots of size around the localization length $N$ (number of fermion modes $N^2$), the spreading of the fermion should

\begin{figure}[tbp]
\begin{center}
\includegraphics[width=3.4in]{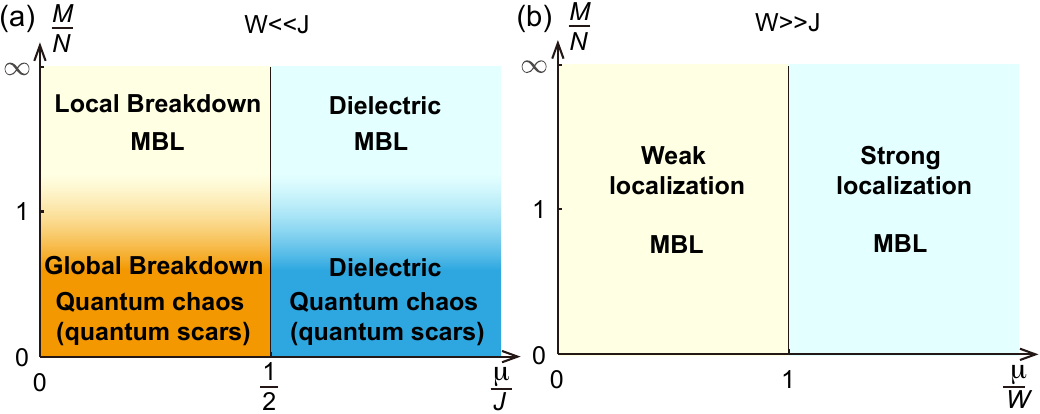}
\end{center}
\caption{The conjectured phase diagram of the quantum breakdown model at large $N$ and $M$, where (a) is at weak disorder potential ($W\ll J$), as a function of $\mu/J$ and $M/N$; and (b) is at strong disorder potential ($W\gg J$), as a function of $\mu/W$ and $M/N$.
}
\label{fig-PD}
\end{figure}

Our conjectures above are summarized into phase diagrams of the quantum breakdown model at large $N$ and $M$, as shown in Fig. \ref{fig-PD}.  At weak disorder potential $W\ll J$, a transition between the dielectric phase and the breakdown phase happens at $\mu/J=1/2$ irrespective of $M/N$. When $M/N\gg1$ (the 1D limit), the system is in a MBL phase, and only local breakdown within order $N$ sites near the initial perturbation can happen when $\mu/J<1/2$. When $M/N\ll1$ (the 3D limit), the system is quantum chaotic (with quantum scar states in certain charge $Q$ sectors), and a global breakdown in the entire space is expected when $\mu/J<1/2$. At strong disorder potential $W\gg J$, the model is persistently in the MBL phase, where the localization is weak when $\mu/W<1$, and strong when $\mu/W>1$.

\section{Discussion}\label{sec:discussion}

We have shown evidences that our quantum breakdown model exhibits a crossover from MBL to quantum chaos at small disorder $W$ as $M/N$ decreases across $1$, and persistent MBL at large disorder $W$. Accordingly, it undergoes a transition from dielectric to breakdown as $\mu/J$ decreases across $1/2$, and shows no breakdown when $W\gg J$, which are summarized in the phase diagrams of Fig. \ref{fig-PD}. At $N=3$ and with disorder potential $W=0$, the model exhibit a Hilbert space fragmentation into exponentially many Krylov subspaces, and is exactly solvable except for very few Krylov subspaces in certain charge $Q$ sectors. The solution indicates features of MBL. Remarkably, the MBL features can occur in the absence of any translational symmetry breaking randomness, namely, when $\mu_m=\mu$, $J^{ijk}_{m,l}=J$ are all uniformly constant and $W=0$. The MBL is robust against a nonzero disorder potential $W>0$. We note that our model show rather distinct features (the local and global breakdown, etc) compared to the interacting Wannier-Stark type models \cite{Nieuwenburg_2019,schulz2019,Moudgalya_2021,herviou2021,sala2020,khemani2020}, although both are describing particles in an electric field. This is because our underlying settings are different: in our setup, the electrons do not feel the electric field until they are ionized from the atoms. For generic $(N,M)$ with $W=0$, we also show that a many-body scar flat band of quantum scar states exists in many charge $Q$ sectors, the origin of which can be understood as the many-body generalization of the line-graph tight-binding models such as the Lieb lattice and kagome lattice models \cite{mielke1991,dumitru2021,lieb1989,bergman2008}. It will be intriguing to generalize this idea to construct more interacting models with many-body flat bands or quantum scar states.

The $N=3$ quantum breakdown model, as an almost exactly solvable MBL model at $W=0$ as we demonstrated, may allow one to further explore analytically the MBL quantum dynamics, conserved quantities and quantum entanglement \cite{serbyn2013,huse2014,chandran2015,ros2015,lian_conserv_2022}, in contrast to many MBL models such as the disordered magnetic field XXZ model \cite{basko2006,gornyi2005,oganesyan2007,marko2008,pal2010} which heavily relies on many-body numerical calculations. Perturbation analysis on the $N=3$ model may give a deeper understanding to the quantum avalanche transition from MBL to thermal states studied recently \cite{roeck2017,thiery2018,luitz2017,goihl2019,crowley2020,morningstar2022,jan2022}. For the $N>3$ models, it may also be possible to find exact or almost exact solutions for certain non-random interactions, which may provide more insights to the phase diagram we conjectured in \cref{fig-PD}. For $N>3$ with random interactions, Hilbert space fragmentation into degenerate Krylov subspaces may still exist, as indicated by the enormous fractal-like degenerate levels other than the many-body scar flat band in the $N=4$ case in \cref{figN>3-W=0}(a).

In the large $N$ limit, with random interactions satisfying \cref{eq-J}, it is possible to study our quantum breakdown model using the large $N$ expansion techniques of the SYK models \cite{sachdev1992fk,polchinski2016xgd,maldacena2016hyu,kitaev2017awl}. It would be particularly interesting to investigate the MBL to chaos crossover and breakdown transition we conjectured in \cref{fig-PD} in this SYK limit. This is, however, beyond the scope of this paper, and will be studied in a separate future paper. Nevertheless, we make a few general observations here. This SYK limit of our model is obviously different from the previous SYK dot lattice models with spatially symmetric interactions (upon random average)  \cite{Berkooz_2017,Gu_2017,davison2017,jian2017,ChenYM_2017}. In our model, one expect the quantum chaos and scrambling to develop towards a fixed direction, in analogy to the chiral SYK models \cite{lian2019,hu2021chiral} and generic chaotic chiral models \cite{hu_integrability_2022,lian2022}, although our model here is not chiral as a lattice model. The conserved charge $Q$ in our model is also very different from the on-site U(1) charge in complex SYK models \cite{gu_complex_2020}. Moreover, one expects a richer phase diagram in the SYK limit of our model. A $1/N$ expansion may be required for investigating the MBL to chaos crossover with respect to $M/N$. We do note that MBL was studied before in SYK models with spatially symmetric interactions \cite{jian2017}.

Since the breakdown phenomenon is the essential mechanism (for converting microscopic signals into macroscopic signals) in many quantum measurement devices such as the Geiger counter, we also anticipate that our quantum breakdown model may provide more insights to the understanding of quantum measurement process. In fact, we noticed that most recently Refs. \cite{hu2022_measurement,liu2022_measurement} has independently studied a spin model in a Bethe tree lattice with similar asymmetric interactions, to simulate the avalanche from a microscopic quantum state into Schr\"odinger cat states in a quantum measurement. Their model also shows Poisson LSS and Hilbert space fragmentation analogous to our model in the MBL limit, and they observed recurrence in the out-of-time-order correlator \cite{liu2022_measurement}.

Our quantum breakdown model further allows a variety of directions of generalizations. First, the number of fermion modes $N$ on each site can be generalized to a site-dependent number $N_m$, which allows one to more freely change the geometry of the ``dielectric gas container". In this case, the formula for the many-body scar flat band degeneracy in \cref{eq:dMm-recursion} simply changes through the substitutions $C_{N}^{n_m}\rightarrow C_{N_m}^{n_m}$ and $h_{n_{m-1},n_m}\rightarrow C_{N_{m-1}}^{n_{m-1}}C_{N_m}^{n_m}$. Secondly, one could assume certain spatial structure within each layer of dielectric gas in Fig. \ref{fig-model}(a), such that the interaction $J_{m,l}^{ijk}$ only acts among nearby fermions, generalizing the model into higher spatial dimensions. An example of such generalized lattices is the Bethe tree studied in \cite{hu2022_measurement,liu2022_measurement}. Moreover, the model can be generalized to a wide class of models of fermions, bosons or spins with different types of asymmetric interactions, which may give rise to a rich variety of non-equilibrium phenomena. It will also be useful to investigate the effect of adding bilinear inter-site fermion hopping terms to our model, allowing a fermion to move without creating additional fermions. Such hopping terms break the conserved charge $Q$, which complicates the model and requires theoretical or numerical methods beyond this work. However, it is possible that such bilinear hopping terms will not destabilize the MBL or the breakdown phases in the presence of disorder $W$. This is because, such a model interpolates between the Anderson localization phase (nonzero hopping and zero interaction) and the MBL phase studied in this paper (zero hopping and nonzero interaction). Another interesting question is whether our Hermitian quantum breakdown model can be related to the non-Hermitian effective quantum models of breakdown \cite{fukui1998,oka2010} in certain limit. We leave the studies of these questions to the future.

%Large $N$, different $N_m$, generalizations, hopping leaking, zero modes, higher dimensions, conserved quantities, chiral systems, MBL without translation breaking, other asymmetric interactions.

\begin{acknowledgments}
The author thank David Huse, Yingfei Gu, Zhenbin Yang, Zhaoyu Han and Elliott Lieb for insightful discussions. The author also thanks the referees for valuable suggestions. 
This work is supported by the Alfred P. Sloan Foundation, the National Science Foundation through Princeton University’s Materials Research Science and Engineering Center DMR-2011750, and the National Science Foundation under award DMR-2141966. Additional support is provided by the Gordon and Betty Moore Foundation through Grant GBMF8685 towards the Princeton theory program.
\end{acknowledgments}

%\newpage
%\ 
%\newpage

\appendix

\section{Reduced entanglement entropy in the rotated basis of the $N=3$ model}\label{app:rot-basis-N=3}

\begin{figure}[tbp]
\begin{center}
\includegraphics[width=3.3in]{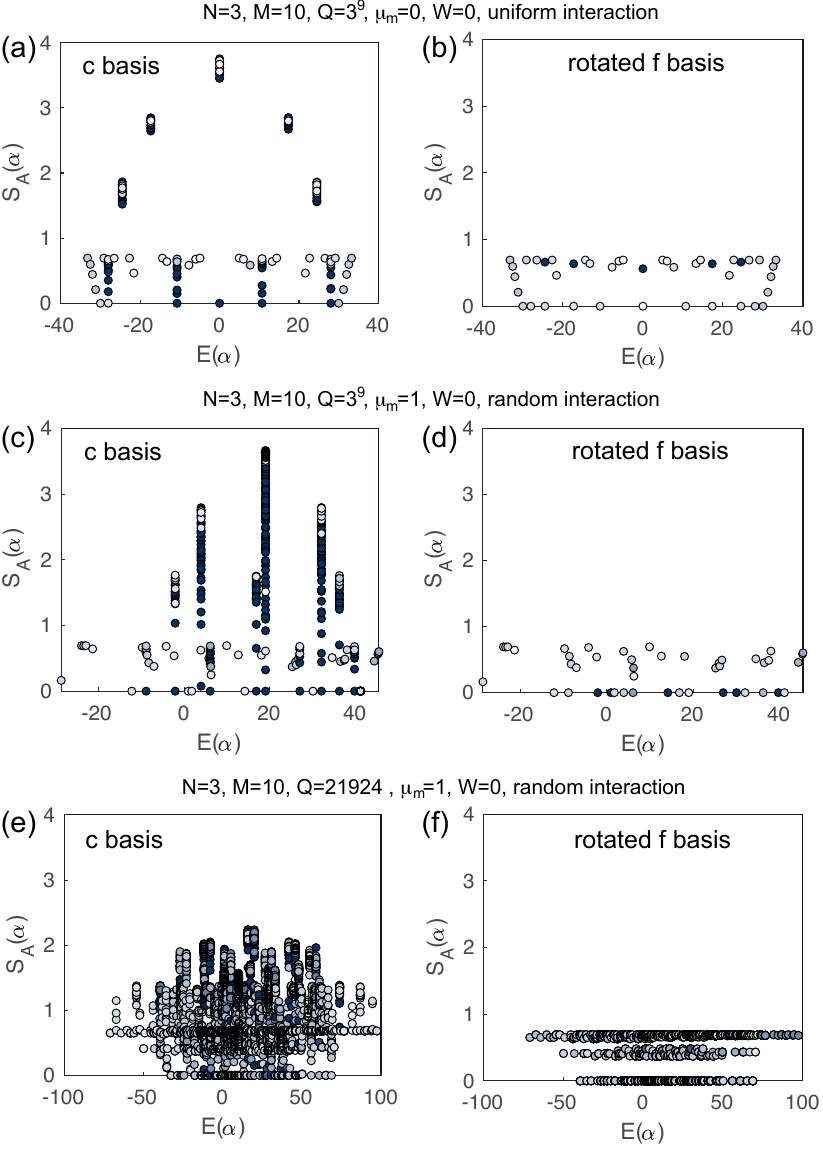}
\end{center}
\caption{For $(N,M)=(3,10)$ at disorder potential $W=0$, the entanglement entropy of the first $\lfloor M/2\rfloor$ sites calculated in the original fermion basis $c_{m,l}$ in \cref{eq-HI} (the left panels) and the rotated basis $f_{m,l}$ in \cref{eq:H3} (the right panels in the same row). The interaction strength is $J=10$. In the rotated basis, a small disorder potential $H_{\text{dis}}=\sum_{m=1}^{M}\sum_i^{N} \nu_{m,i}f^\dag_{m,i}f_{m,i}$ with $\sqrt{\langle \nu_{m,i}^2\rangle}=10^{-6}$ is added, to pin the eigenstates into the Krylov subspace basis we discussed in \cref{sec-N=3-W=0}. The parameters of the first, second, third rows are the same as those of the first, second, third rows in \cref{figN=3-W=0}, respectively.
}
\label{figNe3-We0_rot}
\end{figure}

For the $N=3$ model at disorder potential $W=0$, \cref{figNe3-We0_rot} shows a comparison between the entanglement entropy from ED calculation in the original fermion basis $c_{m,l}$ in \cref{eq-HI} and that in the rotated basis $f_{m,l}$ in \cref{eq:H3} (where only the first fermion mode can move rightward via interaction). The parameters are the same as those in \cref{figN=3-W=0}. In the rotated $f_{m,l}$ basis, a disorder potential $H_{\text{dis}}=\sum_{m=1}^{M}\sum_i^{N} \nu_{m,i}f^\dag_{m,i}f_{m,i}$ with $\sqrt{\langle \nu_{m,i}^2\rangle}=10^{-6}$ is added. As a result, the eigenstates in the rotated $f_{m,l}$ basis are pinned within each Krylov subspace we discussed in \cref{sec-N=3-W=0}, thus all the states show low entanglement entropy no larger than $\ln 2$. This reflects the fact that the model in each Krylov subspace is equivalent to a single particle tight-binding model, which has entanglement entropy at most $\ln 2$. This also implies that the seemingly high entanglement entropy in the original $c_{m,l}$ basis is because the ED arbitrarily chooses a high-entanglement basis within each degenerate subspace, which are superposition of different degenerate Krylov subspaces.

\section{The interaction between subsystems in arbitrary charge $Q$ sectors of the $N=3$ model}\label{app:any-charge-sector-N=3}

This appendix is to supplement our discussion in \cref{sec:N=3-generic-Q-approx}, to write down more explicitly the interaction between neighboring subsystems in certain Krylov subspaces of an arbitrary charge $Q$ sector of the $N=3$ model with $W=0$ given in \cref{eq:H3}.

Following the notations defined in and below \cref{eq:Qgeneral}, each charge $Q$ uniquely corresponds to a reference configuration $\{n_m^0\}$ with $0\le n_m^0\le 2$, and $m_{f,p}$ is the $p$-th site with $n_m^0>0$ ($1\le p\le M_Q$). Generalizing \cref{eq:sub-Hil-3-mb}, we define the following Fock state basis $|\{m_{s,p}\},\{m_{b,p}\},\{s_m^{j}\}\rangle$ in a generic charge $Q$ sector:
%\begin{equation}\label{eq:sub-Hil-3-mbp1}
%\begin{split}
%&|\{m_{s}^p,m_{b}^p,s_m^{j}\}\rangle =\otimes_{p=1}^{M_Q}|m_{s}^p,m_{b}^p,\{s_m^{j}\}\rangle_p,
%\end{split}
%\end{equation}
\begin{equation}\label{eq:sub-Hil-3-mbp}
\begin{split}
&|\{m_{s,p}\},\{m_{b,p}\},\{s_m^{j}\}\rangle \\
&=\prod_{p=1}^{M_Q}\Big(f_{m_{s,p},s^1_{m_{s,p}}}^\dag\prod_{m=m_{b,p}+1}^{m_{s,p}}f^\dag_{m,2}f^\dag_{m,3} \\
&\times \prod_{m=m_{f,p}+1}^{m_{b,p}}f^\dag_{m,s^1_m}f^\dag_{m,s^2_m} \prod_{j=2}^{n_{m_{f,p}}} f^\dag_{m_{f,p},s^j_{m_{f,p}}} \Big)|0\rangle,
\end{split}
\end{equation}
where we restrict the blocking sites $m_{b,p}$ and spreading sites $m_{s,p}$ to satisfy 
\begin{equation}
m_{f,p}\le m_{b,p}\le m_{s,p}\le m_{f,p+1}\ ,
\end{equation}
and we always fix
\begin{equation}
s_{m_{b,p}}^1=1\ ,
\end{equation}
which is necessary for defining $m_b^p$ as a blocking site. In addition, for the state we defined in \cref{eq:sub-Hil-3-mbp} to be nonzero, we require $s_{m_{s,p}}^1=1$ whenever $m_{s,p}>m_{f,p}$, and $s^2_{m^{\text{nz}}_{p}}=2$ or $3$. All the other indices $s_m^j$ can generically take values $1,2$ or $3$ (distinct for the same $m$).

In this representation, each Krylov subspace is labeled by $\{m_{b,p}\},\{s_m^{j}\}$ for a given charge $Q$ sector (i.e., given $\{m_{f,p}\}$). The action of the Hamiltonian in each Krylov subspace can be generically written down as
\begin{equation}
\begin{split}
&H|\{m_{s,p}\}, \{m_{b,p}\},\{s_m^{j}\}\rangle \\
&=(H_{\text{hop}}+H_{\text{b-int}})|\{m_{s,p}\}, \{m_{b,p}\},\{s_m^{j}\}\rangle\ ,
\end{split}
\end{equation}
where $H_{\text{hop}}$ is a part that can be viewed as hoppings of a non-interacting tight-binding model:
\begin{widetext}
\begin{equation}\label{seq:N=3-Hhop}
\begin{split}
&H_{\text{hop}}|\{m_{s,p}\}, \{m_{b,p}\},\{s_m^{j}\}\rangle =\sum_{p'=1}^{M_Q}\Big[ \delta_{s^1_{m_{s,p'}},1}\Big( \vartheta_{m_{s,p'}-m_{b,p'}-1+\delta_{m_{b,p'},m_{f,p'}}}J_{m_{s,p'}-1}|\{m_{s,p}-\delta_{p,p'}\},\{m_{b,p}\},\{s_m^{j}\}\rangle \\
&\qquad\qquad\qquad +\vartheta_{m_{f,p'+1}-m_{s,p'}-1}J_{m_{s,p'}}|\{m_{s,p}+\delta_{pp'}\},\{m_{b,p}\},\{s_m^{j}\}\rangle \Big)+V_{m_{s,p'}}|\{m_{s,p}\}, \{m_{b,p}\},\{s_m^{j}\}\rangle \Big]\ , 
\end{split}
\end{equation}
where $\vartheta_m$ is the integer variable Heaviside function defined in \cref{eq-heaviside}, $J_m$ is defined in \cref{eq-Jm}, and $V_{m}=\sum_{m=1}^{M}\mu_mn_m$ with $n_m$ being the site $m$ number of fermions of state $|\{m_{s,p}\}, \{m_{b,p}\},\{s_m^{j}\}\rangle$. \cref{seq:N=3-Hhop} is nothing but the generalization of \cref{eq:H-tb-mb}, namely, an effective tight-binding Hamiltonian for some effective particles on sites $m_{s,p}$ within subregions $m_{f,p}\le m_{s,p}<m_{f,p+1}$. The term $H_{\text{b-int}}$ is a boundary interaction term
\begin{equation}\label{seq:N=3-Hb-int}
\begin{split}
&H_{\text{b-int}}|\{m_{s,p}\}, \{m_{b,p}\},\{s_m^{j}\}\rangle =\sum_{p'=2}^{M_Q}\Big( \delta_{m_{s,p'-1},m_{f,p'}} \delta_{n_{m_{f,p'}},3}J_{m_{f,p'}-1}|\{m_{s,p}-\delta_{p,p'-1}\},\{m_{b,p}\},\{s_m^{j}\}\rangle \\
&+\delta_{m_{s,p'-1},m_{f,p'}-1} \delta_{n_{m_{f,p'}},0}J_{m_{f,p'}-1}|\{m_{s,p}+\delta_{p,p'-1}\},\{m_{b,p}\},\{s_m^{j}\}\rangle\Big)\ ,
\end{split}
\end{equation}
\end{widetext}
where $n_{m_{f,p'}}$ in the delta functions $\delta_{n_{m_{f,p'}},3}$ and $\delta_{n_{m_{f,p'}},0}$ is the number of fermions on site $m_{f,p'}$ of the state $|\{m_{s,p}\}, \{m_{b,p}\},\{s_m^{j}\}\rangle$. This dependence on the number of fermions $n_{m_{f,p'}}$ makes \cref{seq:N=3-Hb-int} an interaction term for the effective particles located at sites $m_{s,p}$, although it is confined to the boundary sites $m_{f,p'}$. Note that if $m_{b,p'}>m_{f,p'}$, the fermion number $n_{m_{f,p'}}$ cannot be changed by fermions on the right of site $m_{f,p'}$, and one will always have $\delta_{n_{m_{f,p'}},3}=\delta_{m_{s,p'-1},m_{f,p'}} $, and $\delta_{n_{m_{f,p'}},0}=\delta_{m_{s,p'-1},m_{f,p'}-1} $, in which case \cref{seq:N=3-Hb-int} becomes a free term for the effective particles  on sites $m_{s,p}$ as an additional part of \cref{seq:N=3-Hhop}. This interaction between the $(p'-1)$-th and $p'$-th subsystem is only present when $n_{f,p'}^0=1$ and the Krylov subspace has $m_{b,p'}=m_{f,p'}$, making it a very weak interaction. This interaction is a hardcore repulsion on site $m_{f,p'}$ between the two effective particles on sites $m_{s,p'-1}$ and $m_{s,p'}$, as is obvious from the fact that $m_{s,p'-1}=m_{s,p'}$ is forbidden.

\begin{figure*}[tbp]
\begin{center}
\includegraphics[width=6.8in]{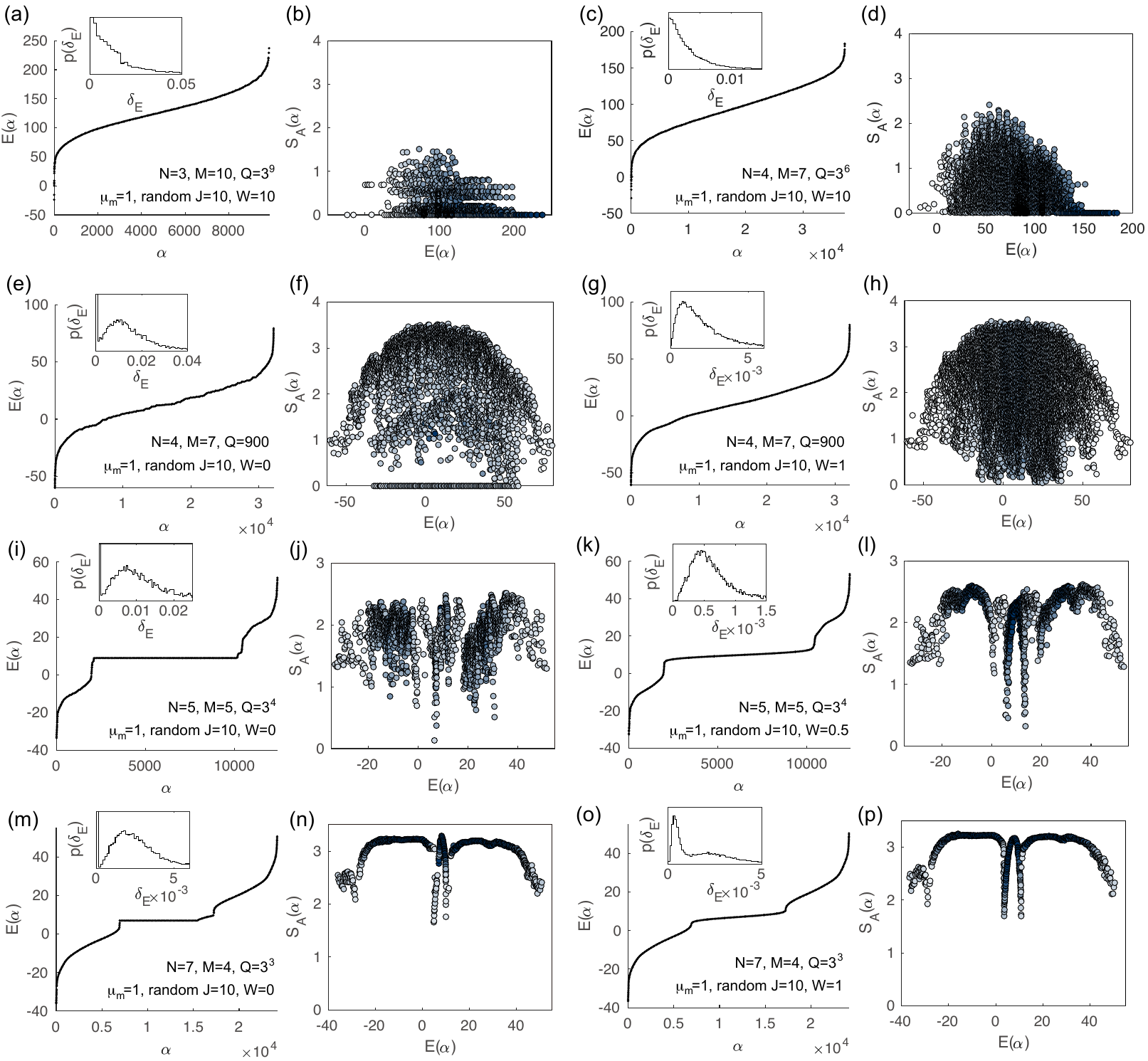}
\end{center}
\caption{ED calculation of more charge $Q$ sectors of the models with $N=3,4,5$ and $7$ and the largest $M$ calculable. The parameters are labeled in the panels, where random interaction $J=10$, the chemical potential is uniformly $\mu_m=\mu=1$. The first and third columns are the energy levels $E(\alpha)$ and their LSS in the insets. The second and fourth columns show the eigenstate entanglement entropy in subregion $A$ (the first $\lfloor M/2\rfloor$ sites) for the energy spectra on their left (the darker the higher density of dots). The disorder potential is $W=10$ in (a)-(d), to show that the MBL behaviors in the $M/N>1$ case remain robust when $W$ is comparable to $J$. We set $W=0$ in panels (e), (f), (i), (j), (m), (n), and $W=1$ or $0.5$ in panels (g), (h), (k), (l), (o), (p).
}
\label{figS-N>3}
\end{figure*}

\section{Additional ED results for $N>3$ models}\label{app:N>3}

This appendix shows more examples of ED calculations in different charge $Q$ sectors of the $N>3$ models.

%\begin{onecolumngrid}
%\end{onecolumngrid}

\cref{figS-N>3} show additional charge $Q$ sectors of the models with $N=3,4,5$ and $7$ and the largest $M$ calculable, as a supplement to \cref{figN=3-W=0,figN=3-W>0,figN>3-W=0,figN>3-W>0} in the main text (the same system sizes). The detailed descriptions and parameters are given in the caption. The results show the same trend as the main text figures show, that the system tends to MBL if $M/N>1$ (with Poisson or semi-Poisson LSS and many eigenstates with area law entanglement entropy, subject to finite system sizes), and behaves quantum chaotically when $M/N<1$ (with Wigner-Dyson LSS and most eigenstates with volume law entanglement entropy).

\begin{figure}[tbp]
\begin{center}
\includegraphics[width=3.3in]{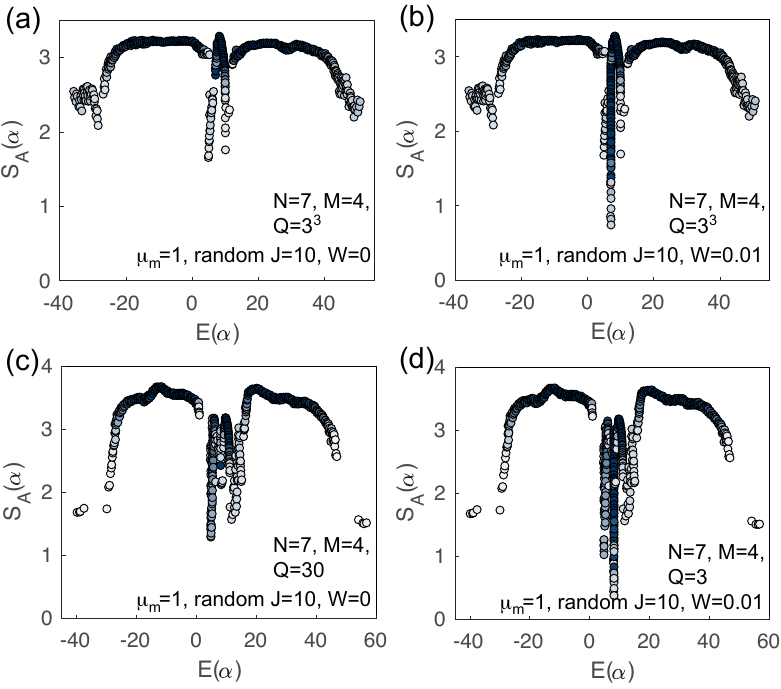}
\end{center}
\caption{Changes of the eigenstate entanglement entropy of the charge $Q=3^3$ (panels (a)-(b)) and $Q=30$ (panels (c)-(d)) sectors of the $(N,M)=(7,4)$ model upon adding a small disorder potential $W=0.01$ (the right two panels). The parameters are given in the panels. The many-body scar flat band is at energy $E=7$ in the charge $Q=3^3$ sector, and is at energy $E=8$ in the $Q=30$ sector.
}
\label{figS-N>3-smallW}
\end{figure}

\begin{figure}[tbp]
\begin{center}
\includegraphics[width=3.3in]{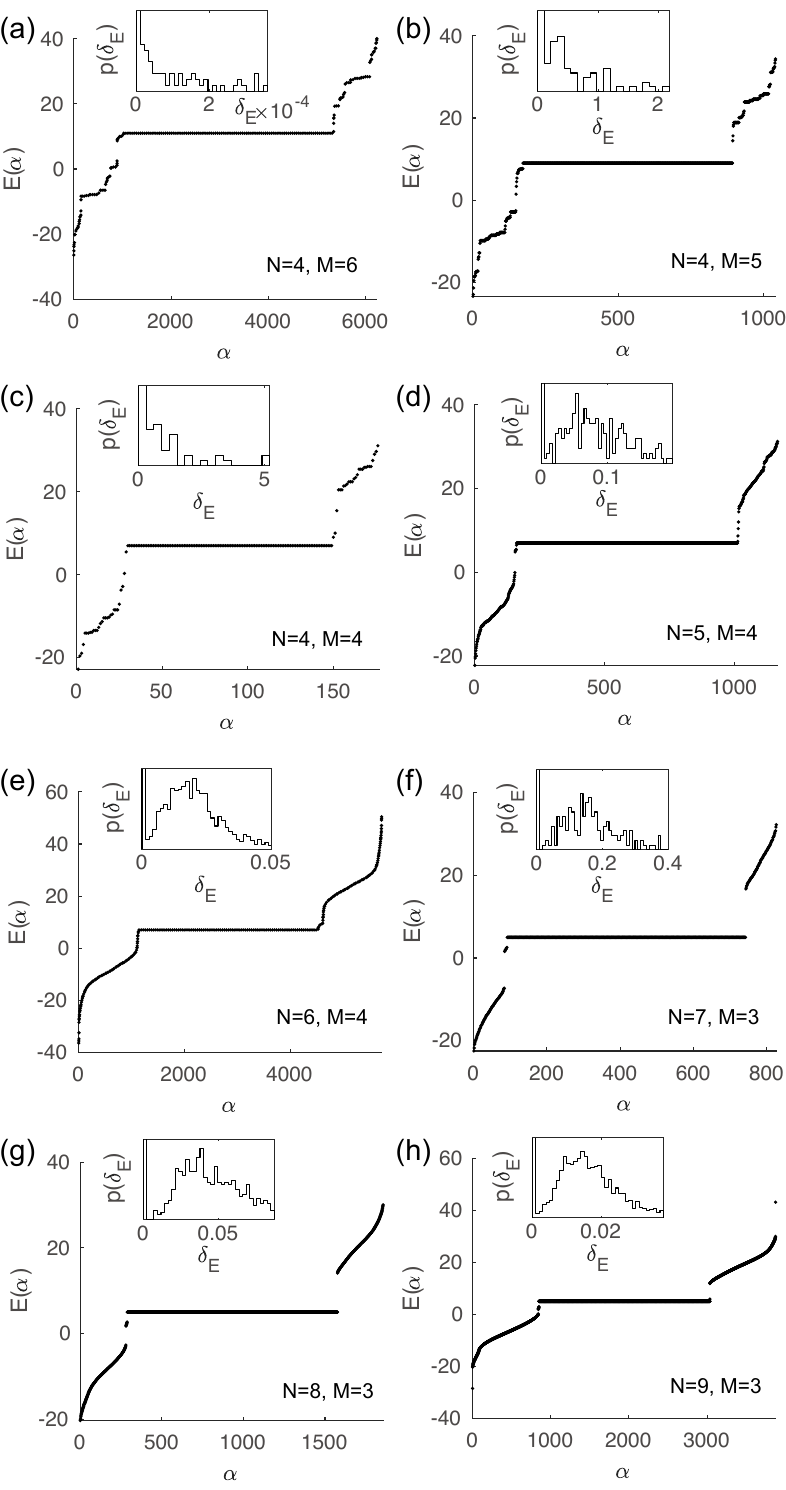}
\end{center}
\caption{The energy spectrum $E(\alpha)$ (from low to high) and LSS (the insets) of various small system sizes $(N,M)$ (see values in the panels, with $NM\approx 16\sim 27$) with $W=0$, where the charge sector is always $Q=3^{M-1}$, the interaction is random with interaction strength $J=10$, and the chemical potential is uniformly $\mu_m=\mu=1$. The LSS shows a clear crossover from Poisson to Wigner-Dyson as $M/N$ decreases across $1$.
}
\label{figS-smallNM}
\end{figure}

The many-body scar flat band in a charge $Q$ sector can be immediately fixed into a localized eigenbasis when a small disorder potential $W$ is added, which breaks the degeneracy. \cref{figS-N>3-smallW} shows two examples of the eigenstate entanglement entropies in the charge $Q=3^3$ and $Q=30$ sectors in the $(N,M)=(7,4)$ model. When a disorder potential $W=0.01$ is added to the system, while the entanglement entropies of most eigenstates are almost unchanged, the entanglement entropy of the many-body scar flat band in the middle immediately drops significantly, indicating the real space localization of the eigenbasis.

To test the MBL to chaos behavior crossover around $M/N\approx1$, we further show in \cref{figS-smallNM} the energy spectra and LSS of smaller system sizes $(N,M)$ at $W=0$, where we always choose the $Q=3^{M-1}$ sector as an example. The interaction is random with a strength $J$, and chemical potential is uniformly $\mu_m=\mu=1$. One can clearly see a crossover from Poisson to Wigner-Dyson LSS as $M/N$ decreases across $1$, in consistency with a MBL to chaos behavior crossover.

\section{More on the classical degeneracy of many-body scar flat band for $M=2$ sites}\label{app:zero-modes}

Here for the model with $M=2$ sites, we give a theoretical quantitative understanding of the dependence of the classical degeneracy $d_2^0$ of many-body scar flat band on $N$ as we showed in \cref{figM=2}(d)-(f). In particular, a transition is observed between $N=28$ and $N=29$ in \cref{figM=2}: for $N\le 28$, most charge $Q$ sectors have a many-body scar flat band, while for $N\ge 29$, only charge sectors with $Q$ close to $0$ and $4N$ have a many-body scar flat band. We explain these two regimes in two subsections below, respectively.

By \cref{eq:d20}, classical degeneracy $d_2^0(n_1,n_2)$ is given by
\begin{equation}\label{seq:d20}
d_2^0(n_1,n_2)=h_{n_1,n_2}-h_{n_1+1,n_2-3}-h_{n_1-1,n_2+3}\ ,
\end{equation}
where $h_{n_1,n_2}=C_{N}^{n_1}C_{N}^{n_2}$. We first note that for $M=2$, the number of fermions $n_2$ on site $m=2$ can be represented by charge $Q$ and the number of fermions $n_1$ on site $m=1$ as
\begin{equation}
n_2=Q-3n_1\ .
\end{equation}
Therefore, we can express $h_{n_1,n_2}$ as a function of $Q$ and $n_1$ as
\begin{equation}
\begin{split}
&h_{n_1,n_2}=h_{n_1,Q-3n_1}=C_{N}^{n_1}C_{N}^{Q-3n_1} \\
&=\frac{N!^2}{n_1!(N-n_1)!(Q-3n_1)!(N-Q+3n_1)!}\ .
\end{split}
\end{equation}
We now examine the condition for a positive classical degeneracy $d_2^0>0$.

\subsection{Small $N$}

As an approximation, as long as $N$ is not too small, we can regard $\ln h_{n_1,Q-3n_1}$ as a continuous function of $n_1$. The reason to take a log is because $h_{n_1,Q-3n_1}$ can change by several orders of magnitude as a function of $n_1$. For a fixed charge $Q$, we can approximate the classical degeneracy $d_2^0(n_1,n_2)=d_2^0(n_1,Q-3n_1)$ in \cref{seq:d20} by Taylor expansion as
\begin{equation}\label{seq:d20-estimate0}
\begin{split}
&d_2^0(n_1,Q-3n_1) \approx h_{n_1,Q-3n_1}\\
&\times \left[- \frac{d^2\ln h_{n_1,Q-3n_1}}{d n_1^2} -\left(\frac{d\ln h_{n_1,Q-3n_1}}{d n_1}\right)^2 -1 \right].
\end{split}
\end{equation}
By a straightforward calculation, we find
\begin{equation}
\begin{split}
&\frac{d\ln h_{n_1,Q-3n_1}}{d n_1}\approx \ln (N-n_1)-\ln n_1 \\
&\qquad\qquad  +3\ln (Q-3n_1) -3\ln (N-Q+3n_1)\ ,
\end{split}
\end{equation}
and
\begin{equation}
\begin{split}
&\frac{d^2\ln h_{n_1,Q-3n_1}}{d n_1^2}\approx -\frac{1}{N-n_1}-\frac{1}{n_1}\\
&\qquad \qquad \qquad\qquad  -\frac{9}{Q-3n_1} -\frac{9}{N-Q+3n_1}\ .
\end{split}
\end{equation}

\begin{figure}[tbp]
\begin{center}
\includegraphics[width=3in]{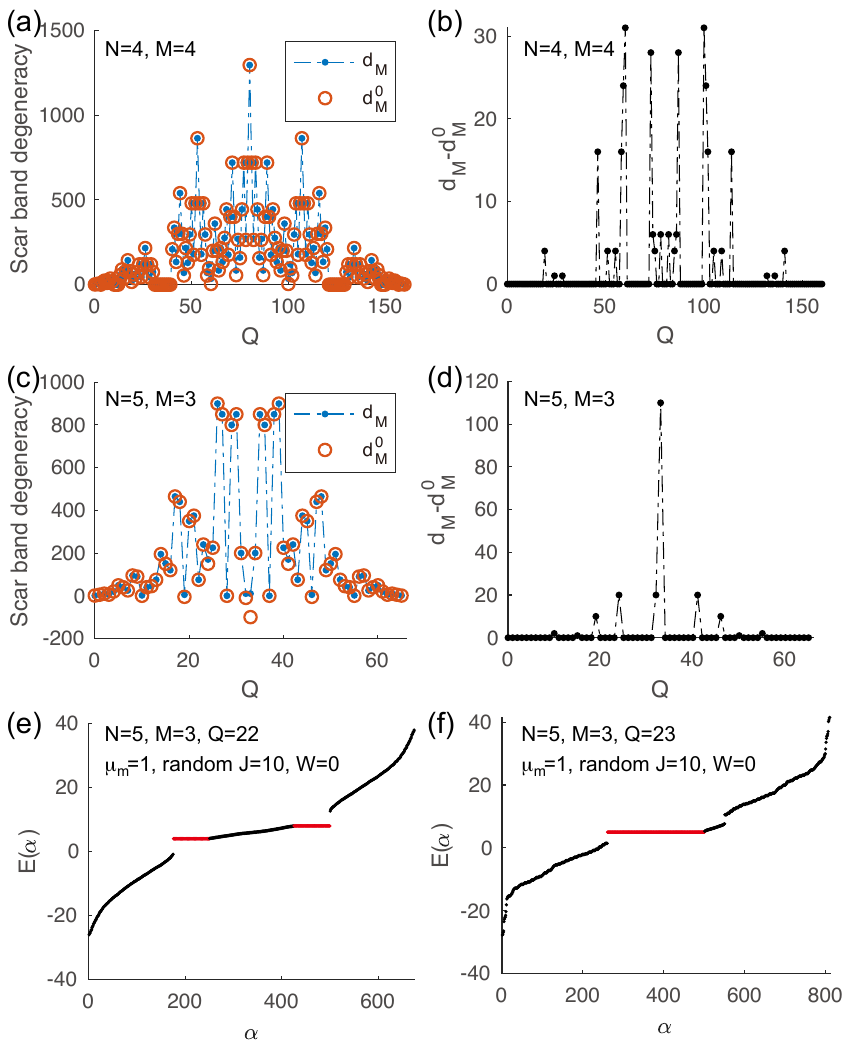}
\end{center}
\caption{(a)-(d) The comparison between the classical degeneracy $d_M^0$ (blue dots) from \cref{eq:dMm0-recursion} and quantum degeneracy $d_M$ (red circles) in ED, for $(N,M)=(4,4)$ and $(N,M)=(5,3)$, respectively. The parameters are $\mu_m=\mu=1$, random interaction strength $J=10$, and $W=0$. (e)-(f) The energy spectrum of two example charge sectors in the $(N,M)=(5,3)$ model, which has two and one many-body scar flat bands (red), respectively.
}
\label{figS-DegeM=34}
\end{figure}

In order to obtain a positive classical degeneracy $d_2^0$ in \cref{seq:d20-estimate0}, we should have $\frac{d^2\ln h_{n_1,Q-3n_1}}{d n_1^2}<0$, and at the same time $\frac{d\ln h_{n_1,Q-3n_1}}{d n_1}$ as close to zero as possible. Numerically, it can be shown that $\frac{d\ln h_{n_1,Q-3n_1}}{d n_1}=0$ when 
\begin{equation}\label{seq:n1-estimate}
n_1\approx \frac{Q}{4}\ ,
\end{equation}
at which one has
\begin{equation}
\frac{d^2\ln h_{n_1,Q-3n_1}}{d n_1^2}\approx -40\left(\frac{1}{Q}+\frac{1}{4N-Q}\right)\ .
\end{equation}
Note that \cref{seq:n1-estimate} implies $n_2=Q-3n_1\approx Q/4$. Therefore, we estimate the largest classical degeneracy $d_2^0$ is reached around the fermion number configuration $n_1\approx n_2\approx Q/4$, estimated by the formula
\begin{equation}\label{seq:d20-estimate1}
\begin{split}
&d_2^0(Q) \approx h_{n_1,Q-3n_1} \left[40\left(\frac{1}{Q}+\frac{1}{4N-Q}\right)-1\right]\ .
\end{split}
\end{equation}

It is easy to see that $d_2^0$ is minimal when $Q=2N$ (note that $0\le Q\le 4N$). Therefore, if $d_2^0>0$ in the $Q=2N$ sector, we would expect almost all (not all because this estimation is very crude) charge $Q$ sectors to have a many-body scar flat band. Since
\begin{equation}\label{seq:d20-estimate2}
\begin{split}
&d_2^0(Q=2N) \approx h_{n_1,Q-3n_1} \left(\frac{40}{N}-1\right)\ ,
\end{split}
\end{equation}
which is positive only if $N<40$, we conclude that the criterion for having many-body flat band in almost all charge $Q$ sectors is $N\lesssim 40$. In the actual numerical calculation in \cref{figM=2}(d)-(f), this criterion is $N\le 28$, which is reasonably not far from our crude estimation.

\subsection{Large $N$}

As we have shown above, for large $N$ ($N\ge 29$ from numerical calculation in \cref{figM=2}(d)-(f)), the many-body scar flat band will not occur in most charge $Q$ sectors. However, here we show that there is always a many-body scar flat band when $Q$ is close enough to $0$ or $4N$. Since the classical degeneracy in \cref{seq:d20} is symmetric between $Q$ and $4N-Q$, we only consider small $Q$ here.

The proof is easy by noting that the configuration $(n_1,n_2)=(0,Q)$ always has a positive classical degeneracy $d_2^0$ at sufficiently small $Q$. By  \cref{seq:d20}, one has
\begin{equation}
\begin{split}
&d_2^{0}(0,Q)=h_{0,Q}-h_{1,Q-3} \\
&=C_N^Q\left[ 1-\frac{NQ(Q-1)(Q-2)}{(N-Q+1)(N-Q+2)(N-Q+3)} \right]\\
&\approx C_N^Q\left( 1-\frac{Q^3}{N^2} \right)\ .
\end{split}
\end{equation}
Therefore, at large $N$, when $Q<N^{2/3}$, there is always a many-body scar flat band with fermion number configuration $(n_1,n_2)=(0,Q)$. Similarly, when $Q>4N-N^{2/3}$, there is always a many-body scar flat band with fermion number configuration $(n_1,n_2)=(N,Q-3N)$. 
%So the condition for having scar flat band with $n_1=0$ is $Q<N^{2/3}$. At large $N$, this is the only case with scar bands.

\section{Additional results for many-body scar flat band with $M>2$ sites}\label{app:M>2}

In \cref{figS-DegeM=34}, we provide additional ED results as a supplementary to the main text \cref{figM=3}. \cref{figS-DegeM=34}(a)-(d) show the comparison between the quantum degeneracy $d_M$ and classical degeneracy $d_M^{0}$ (calculated by \cref{eq:dMm0-recursion}) of many-body scar flat band in all the charge $Q$ sectors, where the system sizes are $(N,M)=(4,4)$ and $(N,M)=(5,3)$, respectively. Again, the difference between $d_M$ and $d_M^{0}$ is zero in most charge $Q$ sectors, and small compared to $d_M$ in all the charge $Q$ sectors. 

In addition, we show a very rare charge $Q$ sector example in \cref{figS-DegeM=34}(e) which has two many-body scar flat bands, in contrast to most charge $Q$ sectors which has only one many-body scar flat band (e.g., \cref{figS-DegeM=34}(f)).

\bibliography{qa_ref}

\end{document}